\documentclass[aps, prd, reprint, longbibliography, nofootinbib,superscriptaddress, floatfix]{revtex4-2}
\pdfoutput=1

\usepackage{amsmath,amssymb,amsfonts}
\usepackage{mathrsfs}
\usepackage{bbm}
\usepackage{slashed}
\usepackage{graphicx}
\usepackage{comment}
\usepackage{verbatim}
\usepackage[T1]{fontenc} 
\usepackage[utf8]{inputenc}
\usepackage[colorlinks]{hyperref}
\usepackage{mathbbol}
\usepackage[dvipsnames]{xcolor}
\usepackage{svg}
\usepackage{flushend}
\usepackage{physics}
\usepackage{multirow}
\usepackage{graphicx}
\usepackage{dcolumn}
\usepackage{bm}
\usepackage{tensor}
\usepackage{comment}
\usepackage[utf8]{inputenc} 
\usepackage{graphicx,color,overpic,mathtools}
\usepackage{amsthm,amsmath,amssymb,hyperref,mathrsfs}
\usepackage{braket,bm,bbm}
\usepackage{cancel}
\usepackage{physics}
\usepackage{float}
\usepackage[english]{babel}
\usepackage{graphicx}
\hypersetup{
    colorlinks=false,
    pdfborder={0 0 0},
}

\allowdisplaybreaks

\usepackage[normalem]{ulem}

\newcommand{\qu}[1]{``{#1}''} 

\begin{document}

\title{Asymptotic safety is not generic}

\author{Francesco Del Porro}
\email[]{francesco.del.porro@nbi.ku.dk}
\affiliation{Center of Gravity, Niels Bohr Institute, Blegdamsvej 17, DK-2100 Copenhagen \O, Denmark}
\affiliation{Niels Bohr International Academy, Niels Bohr Institute, Blegdamsvej 17, DK-2100 Copenhagen \O, Denmark}

\author{Francesco Ferrarin}
\email[]{francesco.ferrarin@nbi.ku.dk}
\affiliation{Center of Gravity, Niels Bohr Institute, Blegdamsvej 17, DK-2100 Copenhagen \O, Denmark}
\affiliation{Niels Bohr International Academy, Niels Bohr Institute, Blegdamsvej 17, DK-2100 Copenhagen \O, Denmark}

\author{Alessia Platania}
\email[]{alessia.platania@nbi.ku.dk}
\affiliation{Center of Gravity, Niels Bohr Institute, Blegdamsvej 17, DK-2100 Copenhagen \O, Denmark}
\affiliation{Niels Bohr International Academy, Niels Bohr Institute, Blegdamsvej 17, DK-2100 Copenhagen \O, Denmark}
\affiliation{Institute of Physics, University of Graz,
Universit\"atsplatz 5, 8010 Graz, Austria}

\begin{abstract}
Asymptotically safe quantum gravity proposes a non-perturbative ultraviolet completion of gravity in which the high-energy behavior is controlled by an ultraviolet fixed point---the Reuter fixed point---for which there is robust evidence from a wide range of settings and truncations. The existence of asymptotic safety requires non-trivial zeros of beta functions, raising the question of how generic such fixed-point solutions are. In this work, we tackle this issue by investigating chiral perturbation theories.
We focus on SU(2) and SU(3) chiral models, which share key features with gravity such as perturbative non-renormalizability or nonlinear interactions, but which are known to admit a different ultraviolet completion via new degrees of freedom. Using the functional renormalization group, we study these theories within a quartic-derivative truncation and find no evidence for an interacting fixed point. This result suggests that asymptotic safety is not generic, but rather a feature pertaining to certain special systems.
\end{abstract}

\maketitle

\section{Introduction}

Gravity is perturbatively non-renormalizable: the Newton's coupling has negative mass dimension, so quantum corrections generate an infinite tower of counterterms and the theory loses predictivity at high energies when treated as a standard perturbative quantum field theory. Asymptotic safety (AS)~\cite{Percacci:2017fkn,Reuter:2019byg,Bonanno:2020bil,Pawlowski:2020qer,Knorr:2022dsx, Eichhorn:2022gku, Morris:2022btf, Martini:2022sll, Wetterich:2022ncl, Platania:2023srt, Saueressig:2023irs, Pawlowski:2023gym, Bonanno:2024xne,Eichhorn:2026uqj} offers a way out within the framework of quantum field theory itself. If the renormalization group (RG) flow admits an interacting ultraviolet (UV) fixed point with a finite number of relevant directions, the theory is predictive and well-defined at all scales, despite being non-renormalizable in the perturbative sense. For gravity, such a fixed point---the Reuter fixed point---has been found with the functional renormalization group (FRG)~\cite{Dupuis:2020fhh} in a wide range of settings and truncations, including increasingly refined approximations of the effective average action, different parametrizations and regulators, and various matter and higher-curvature extensions~\cite{Reuter:2001ag, Lauscher:2002sq, Codello:2006in,Benedetti:2009rx,Dona:2013qba, Gies:2015tca, Gies:2016con,Biemans:2016rvp,Biemans:2017zca, Christiansen:2017cxa, Hamada:2017rvn, Knorr:2017fus, Christiansen:2017bsy, Falls:2017lst,Knorr:2018fdu,Eichhorn:2019ybe, Alkofer:2020vtb,Falls:2020qhj, Knorr:2021lll, Knorr:2021slg, Kluth:2022vnq, Pastor-Gutierrez:2022nki,Baldazzi:2023pep,Eichhorn:2025sux}. Independently, lattice computations including Euclidean and causal dynamical triangulations can identify a continuous phase transition~\cite{Ambjorn:2024qoe,Ambjorn:2024bud,Loll:2025eks}, which in statistical physics is described by non-trivial fixed points. All in all, the evidence for the Reuter fixed point is robust.

The existence of AS requires non-trivial zeros of the beta functions, which raises a natural question: \textit{how generic are such fixed-point solutions?} If an interacting UV fixed point were the typical fate of non-renormalizable theories, AS would be a broad paradigm, with gravity as one instance among many. If instead it were rare, its presence in gravity would point to special features of the gravitational interaction that deserve to be identified. Addressing this question requires a controlled setting outside gravity, with comparable structure but a different known UV behavior.

Chiral perturbation theory~\cite{Weinberg:1978kz} is a natural candidate. As an effective theory of massless mesons, it is organized as a derivative expansion, is perturbatively non-renormalizable, and has nonlinear interactions dictated by the symmetry-breaking pattern. These are all features shared with gravity. Unlike gravity, however, its UV completion is known to proceed through new degrees of freedom: the linear sigma model at intermediate scales and, ultimately, Quantum Chromodynamics. If AS were a generic mechanism, one could expect an interacting fixed point to appear in the chiral models as well, in competition with or in addition to this completion. Related questions have been studied for nonlinear sigma models (NLSMs) in~\cite{Codello:2008qq,Percacci:2009fh,Flore:2012ma}.

In this Letter, we study SU(2) and SU(3) chiral models in the chiral limit within the FRG, using a truncation of the effective action to quartic order in derivatives. We derive the flow equations for the dimensionless couplings and search systematically for zeros of the beta functions. We find no evidence for an interacting fixed point in either model. This holds under variations of the regularization scheme. Our result suggests that AS is not a generic feature of non-renormalizable theories. 

We then go beyond the numerical result and attempt to trace its origin to the structure of the beta functions. Comparing the chiral models with gravity, in particular with a quadratic truncation in standard and unimodular settings, we point to two main differences. The first is the presence of a volume term in gravity, which has no counterpart in the chiral models. The second concerns negative contributions from the Faddeev–Popov ghosts, which are clearly absent in the chiral case. These and other differences may point to structural criteria for when a fixed point can arise, and support the view that AS pertains to special systems rather than being a general property of perturbatively non-renormalizable theories.

The paper is organized as follows. In Sect.~\ref{sec:NLSM_FRG} we introduce the chiral models and the FRG setup. Sect.~\ref{sec:NO_FP} analyzes fixed points of the beta functions and the robustness of their inexistence. In Sect.~\ref{sec:NLSM_vs_gravity} we compare with gravity and discuss why AS is not generic. We draw our conclusions in Sect.~\ref{sec:discussion}. Technical details are collected in the appendices.

\section{Chiral NLSM within the FRG} \label{sec:NLSM_FRG}
A NLSM is, from a geometric perspective, a theory of maps between manifolds. The field configurations are smooth functions $\varphi: \mathcal M \to \mathcal N$ from a $D$-dimensional spacetime manifold to an $N$-dimensional target manifold. We consider $\mathcal M$ to be equipped with a flat Euclidean metric, while $\mathcal N$ carries a Riemannian metric $h$. The components $\varphi^a$ may therefore be interpreted both as scalar fields on spacetime and as local coordinates on the target manifold. Because the target-space metric depends non-trivially on the fields, $h_{ab} = h_{ab}(\varphi)$, the theory is intrinsically nonlinear: the fields interact through the geometry of the target manifold itself. Covariant derivatives acting on target-space tensors are defined using the pullback of the spacetime Levi-Civita connection via the embedding~$\varphi$. 

\subsection{Higher-derivative truncations for the NLSM}

The action for the NLSM can be organized in a derivative expansion. The simplest truncation then contains two derivatives and is given by~\cite{Weinberg:1978kz,Howe:1986vm,Hasenfratz:1988rf}
\begin{equation}
    S_{\rm LO}=\frac{1}{2 F} \int  {\rm d}^Dx \, h_{ab}(\varphi) \partial_{\mu}\varphi^a\partial^{\mu}\varphi^b \,, \label{eq:NLSM_LO}
\end{equation}
where $F$ is a coupling constant, akin to the Newton's coupling in the Einstein-Hilbert truncation of a gravitational effective field theory (EFT). Since the fields $\varphi^a$ are coordinates on the target manifold, they are dimensionless, implying that $F$ has mass dimension $[F]=2-D$. Due to the non-trivial geometry of $\mathcal M$, the action~\eqref{eq:NLSM_LO} entails infinitely many interaction terms, encoded in the $\varphi$-dependence of the metric $h_{ab}$. The functional form of $S_{\rm LO}$ is independent of the details of the target manifold, since a single two-derivative operator is allowed.

An important class of NLSMs arises when the target manifold is a compact Lie group $G$. In that case, Eq.~\eqref{eq:NLSM_LO} corresponds to the leading order term of a chiral EFT invariant under a global $G_L \times G_R$ symmetry~\cite{Weinberg:1978kz}
\begin{equation}
    S_{\rm \chi PT}=\frac{1}{F} \int {\rm d}^Dx {\rm Tr}[\partial_{\mu} U^{\dagger} \partial^{\mu}U] \,. \label{eq:chiEFT_LO}
\end{equation} 
Here, $U \in G$ is invariant under the transformation $g_L^{-1}Ug_R$, where $g_L \in G_L$ and $g_R \in G_R$. The map between Eq.~\eqref{eq:NLSM_LO} and Eq.~\eqref{eq:chiEFT_LO} is given by the exponential map $U=\text{exp}(i \varphi^a T_a)$, with $T_a$ being the generators of the Lie algebra of $G$, normalized as $2{\rm Tr}(T_aT_b)=\delta_{ab}$.

Throughout this work, unless specified otherwise, we will consider $D=4$. In this case, the theories associated with $G=SU(2)$ and $G=SU(3)$ describe the dynamics of massless pions and mesons, respectively. Remarkably, in both cases, the NLSM shows an interacting UV fixed point within the truncation~\eqref{eq:chiEFT_LO}, as shown in~\cite{Codello:2008qq}. 

At the next-to-leading order in a derivative expansion, the exponential map $U=\text{exp}(i \varphi^a T_a)$ still allows us to rewrite the chiral EFT as a NLSM. However, the two cases correspond to different Lagrangians, due to the different geometrical structure of the Lie manifolds. In fact, $SU(2) \simeq S^3$ is isomorphic to a 3-sphere, that is, a maximally symmetric spacetime~\cite{Percacci:2009fh}. In this case, the Riemann tensor is completely determined by the metric and the Ricci scalar is constant---see App.~\ref{Appendix:0}. This geometrical constraint reduces the number operator for $G=SU(2)$ down to two four-derivative operators~\cite{Weinberg:1978kz,Gasser:1983yg,Hasenfratz:1988rf,Percacci:2009fh}. The next-to-leading part of the NLSM action is thus
\begin{align} \label{eq:chiEFT_NLO_SU2}
    S_{\rm NLO}^{\rm SU(2)}=-\frac{1}{2F} &\int {\rm d}^4x   \biggl(  \frac{L_1}{2}(h_{ab} \partial_\mu \varphi^a \partial^\mu \varphi^b)^2\\
   & + \frac{L_2}{2}h_{ab} h_{cd}\partial_\mu \varphi^a \partial^\mu \varphi^c  \partial_\nu \varphi^b \partial^\nu \varphi^d \biggr)\,. \nonumber
\end{align} 
Note the difference with~\cite{Flore:2012ma}: within a parity-invariant NLSM up to four derivative, a complete list of operators should also include the operator $h_{ab} \square \varphi^a \square \varphi^b$, with $\Box= \nabla_\mu \nabla^\mu$. However, this operator is removable through field redefinitions~\cite{Weinberg:1978kz,Gasser:1983yg} and it is not usually included in the chiral EFT (see e.g.~\cite{Guerrieri:2020bto}). As we will show later, this term is not automatically generated by the RG flow, unless it is added by hand from the beginning. In other words, this operator can be consistently neglected. For these reasons, we shall not consider it in the present work. In particular, since our work solely focuses on the existence of fixed points for the chiral beta functions, even if this operator were to contribute, it would not influence our analysis. This is because its dimensionless coupling becomes asymptotically free~\cite{Flore:2012ma}, thus not contributing to the flow when searching for a UV fixed point.

For $G=SU(3)$, a third, independent operator appears at the same order in derivatives. Here, the target manifold is not maximally symmetric and the Riemann tensor is determined by the structure constants of the Lie algebra~\cite{Hasenfratz:1988rf,Percacci:2009fh}---see App.~\ref{Appendix:0}. In practice, the operator
\begin{align} \label{eq:chiEFT_NLO_SU3}
   S_{\rm NLO}^{\rm SU(3)} \supset -\frac{L_3}{4F} \int {\rm d}^4x   R_{a b c d} \partial_{\mu} \varphi^{a}\partial_{\nu} \varphi^{b}\partial^{\mu} \varphi^{c}\partial^{\nu} \varphi^{d} 
\end{align}
cannot be reabsorbed in those of Eq.~\eqref{eq:chiEFT_NLO_SU2}. 

\subsection{FRG setup}

To investigate the existence of fixed points in the chiral NLSMs introduced above, we first need to derive the relevant beta functions. To this end, we introduce an RG scale $k$, and define the dimensionless version of the couplings,
\begin{equation}\label{eq:dimlesscouplings}
    f(k):=k^2 F(k) \,, \quad \ell_i(k):=k^2L_i(k) \,,
\end{equation}
with $i=1,2,3$. To study the variation of these couplings along the RG flow, we shall use the FRG~\cite{Dupuis:2020fhh}, as is done in the context of AS in quantum gravity. The FRG is non-perturbative in the interaction couplings, and is thereby suitable to investigate the existence of interacting fixed points of the flow. Once the interaction constants are promoted to running couplings, the variation of the resulting RG-scale-dependent effective action $\Gamma_k$ is governed by the Wetterich equation~\cite{Wetterich:1992yh}
\begin{equation}
    k \partial_k \Gamma_k=\frac{1}{2} \text{STr} \biggl [ \left ( \Gamma_k^{(2)} + \mathcal{R}_k \right )^{-1} k\partial_k \mathcal{R}_k \biggr]\,. \label{eq:Wetterich_eq}
\end{equation}
Here $\Gamma_k^{(2)}$ indicates the Hessian of $\Gamma_k$, the supertrace \text{STr} denotes an integration over momenta as well as a sum over any internal indices, and the regulator $\mathcal R_k$ implements the Wilsonian shell-by-shell integration of fluctuating modes with $p^2>k^2$. As such, modulo reconstruction problem~\cite{Morris:2015oca,Fraaije:2022uhg}, the limits $k \to \infty$ and $k \to 0$ yield, respectively, the microscopic bare action $S$ and the full effective action $\Gamma$, in which all quantum fluctuations have been integrated out.

To derive the RG flow for the chiral NLSMs above, we shall expand the flow equation~\eqref{eq:Wetterich_eq} within a derivative expansion of $\Gamma_k$, including operators with up to four derivatives. On the left-hand side, this boils down to replacing $\Gamma_k$ with
\begin{equation}\label{eq:chiral-lagrangians}
    \Gamma_k=S_{\text{LO},k}+S_{\text{NLO},k}^G\,,
\end{equation}
with $G=SU(2)$ or $G=SU(3)$, and the NLSM couplings replaced by their $k$-dependent counterparts, according to Eq.~\eqref{eq:dimlesscouplings}. The resulting expression is then evaluated on a generic background. To expand the right-hand side of the flow equation, which involves computing the Hessian $\Gamma_k^{(2)}$, we employ the background field approximation. We select a generic background configuration $\bar \varphi^a(x)$, and we expand in fluctuations about that. As emphasized in~\cite{Codello:2008qq}, the usual linear split $\varphi^a(x)= \bar \varphi^a(x)+ \delta \varphi^a(x)$ is not geometrically meaningful, as it represents a difference of coordinates in the target space rather than a tangent vector. Therefore, we choose a covariant expansion by setting Riemann normal coordinates around $\bar \varphi^a(x)=p \in \mathcal N$. The fluctuations $\xi^a(x)$ are tangent vectors at $p$ and define the \qu{geodesic displacement} from $p$ on $\mathcal N$. The full field is defined via the exponential map
\begin{equation} \label{eq:exp_map}
    \varphi(x) = \exp_{\bar\varphi(x)}\!\left(\xi(x)\right)\,.
\end{equation}
Technically, the expansion can be performed as in~\cite{Howe:1986vm}: we choose a geodesic $\Phi^a(s;x)$ on $\mathcal N$, with $s \in [0,1]$, such that
\begin{equation}
  \bar \varphi^a=  \Phi^a\biggr|_{s=0}\,, \quad \varphi^a=\Phi^a\biggr|_{s=1}\,, \quad \xi ^a=\frac{{\rm d}\Phi^a}{{\rm d}s}\biggr|_{s=0} \,,
\end{equation}
and we expand in powers of $\xi^a$. The computation of the Hessian---omitted here for brevity---can be found in App.~\ref{Appendix:1}. Structurally, we have~\cite{Flore:2012ma}
\begin{equation} \label{eq:Hessian}
 \left( \Gamma^{(2)}_k \right)_{ab}= \frac{1}{F(k)} \left(- h_{ab}(\bar \varphi) \square + \Sigma_{ab}(\bar \varphi) \right)\,,
\end{equation}
with
\begin{align} \label{eq:Sigma}
    \Sigma_{ab}(\bar \varphi)  =   \mathcal B^{\mu \nu}_{ab}(\bar \varphi) \nabla_\mu \nabla_\nu + \mathcal C^\nu_{ab}(\bar \varphi) \nabla_\nu + \mathcal D_{ab}(\bar \varphi)  \,.
\end{align}
The shape of the $\mathcal{B,\, C,\, D}$ tensors is given in App.~\ref{Appendix:1}. From now on we will omit the $\bar \varphi$ dependence. The regulator is chosen to be of the form
\begin{equation} \label{eq:regulator}
 \mathcal R_{k,ab}(-\square)= \frac{1}{F(k)}h_{ab} R_k(- \square)\,.
\end{equation}
To check the robustness of our conclusions in the next section, we shall employ different types of shape functions $r(z/k^2):=k^2R_k(z)$. Setting $y=z/k^2$,
\begin{align}
 & r^{\rm Lit}(y)=(1-y) \Theta(1-y)\,, \quad r^{\rm Sharp}(y)=\Theta(1-y) \,, \nonumber \\
 & r^{\rm Exp}(y;A)= \frac{A y}{e^{y}-1}\,, \quad r^{\rm Exp,2}(y)= e^{-y} \,,\\
 & r^{\rm Fra}(y;n)= \frac{1}{1+y^n} \,. \nonumber
\end{align}\label{eq:shape_functs}
Results will be considered robust if they are independent of the choice of regulator.

The trace of the right-hand side of Eq.~\eqref{eq:Wetterich_eq} is computed via heat kernel techniques. In particular, Eq.~\eqref{eq:Sigma} implies the presence of traces of the form
\begin{align}
{\rm Tr} \left( \nabla_{\mu_1} \dots \nabla_{\mu_n}f_n(\square) \right)  \,.
\end{align}
Off-diagonal traces of this type have been studied in~\cite{Groh:2011dw}; their explicit expressions for the present case are given in App.~\ref{Appendix:2}. We note that the part of $\Sigma_{ab}$ involving $\mathcal C^\mu_{ab}$ does not contribute to our truncation order, similarly to what was found in~\cite{Flore:2012ma}.

Matching the coefficients of the operators on the left- and right-hand sides of Eq.~\eqref{eq:Wetterich_eq} yields the evolution of the individual running couplings, defined by the beta functions
\begin{equation}
    \beta_{f}:= k \frac{{\rm d}f}{{\rm d}k} \,, \quad \beta_{\ell_i}:= k \frac{{\rm d}\ell_i}{{\rm d}k} \,.
\end{equation}
Their expression is reported in App.~\ref{Appendix:4}. In the next section, we shall use these beta functions, varying both the spacetime dimension $D$ and the shape function $r(z/k^2)$, to investigate the existence of interacting fixed points akin to the Reuter fixed point of AS in quantum gravity.

\section{(Absence of) fixed points for SU(2) and SU(3) chiral models} \label{sec:NO_FP}

This section tackles the question that motivated our work: whether the RG flow of chiral NLSMs displays non-trivial fixed points in the UV.
Having derived the RG flow for the chiral Lagrangians~\eqref{eq:chiral-lagrangians}, we now turn our attention to the fixed-point structure. The Gaussian fixed point (GFP), characterized by vanishing dimensionless couplings, corresponds to the free theory and is the simplest realization of scale invariance. Its interacting generalization is called a non-Gaussian fixed point (NGFP), defined by simultaneous zeros of the beta functions, with at least one coupling being non-vanishing. The existence of a NGFP provides the basis for AS in quantum gravity, yet it alone does not guarantee a predictive UV completion: this is dictated by the behavior of the RG flow in its vicinity and its stability properties, which are characterized by the critical exponents $\theta_i$. These are defined as minus the eigenvalues of the stability matrix $S_{ij}=\partial_i\beta_j|_{g_*}$, where $g_*$ denotes the fixed point coordinates for a generic set of couplings $g=\{g_i\}$.
The critical exponents determine how perturbations behave away from the fixed point. In particular, to linear order, a perturbation along an eigendirection scales as $\delta g_i(k)\propto k^{-\theta_i}$. The relevant directions defined by $\mathrm{Re}\,\theta_i>0$ are UV-attractive, whereas those with $\mathrm{Re}\,\theta_i<0$ are UV-repulsive and irrelevant. Requiring trajectories to approach the fixed point in the UV constrains the irrelevant couplings in terms of the relevant ones. Therefore, predictivity requires a finite number of relevant directions, since this ensures that UV-complete trajectories depend only on a finite number of independent parameters.

A general caveat in this discussion is the necessity to truncate: in particular, projecting the RG flow onto a restricted set of operators can generate spurious roots of the beta functions, and some solutions may not correspond to genuine fixed points of the theory. Their existence may be dependent on truncation or choice of regulator, and the robustness of a candidate fixed point must therefore be assessed by checking the stability of the result under variations of scheme or other unphysical choices. The existence of a genuine fixed point should not depend on this. In the present analysis, this motivates searching for fixed points under different choices of regulators, as discussed in the previous section. In particular, we search for simultaneous zeros of the beta functions using the different shape functions defined in Eq.~\eqref{eq:shape_functs}. We employed both \textit{NSolve} for a full zeros search and \textit{FindRoot}. Since the latter is a local numerical method whose convergence depends on the initial conditions, we apply it over a uniform grid of initial guesses spanning a bounded region of coupling space, and systematically enlarge the search region to include larger coupling values.

We find roots of the beta functions for the Litim $r^{\rm Lit}(y)$ and sharp cutoff choices $r^{\rm Sharp}(y)$. However, these solutions are not reproduced for the other regulator shape functions, including the exponential and fractional families for the different values of $n$ considered. For the exponential regulator $r^{\rm Exp}(y;A)$, we scanned the values $A\in \{1,5,10 \}$; for the fractional family $r^{\rm Fra}(y;n)$, we analyzed $n \in \{2,4\}$. This strong regulator dependence suggests that the roots found employing the Litim and sharp regulators may be artifacts of the choice of regulator, and we therefore find no robust evidence for the existence of a NGFP within the truncation considered. This is in sharp contrast with the case of the gravitational Reuter fixed point, which is remarkably robust against changes of regularization scheme, gauge fixing, the inclusion of higher derivative operators, as well as the addition of Standard-Model-like matter fields\footnote{NGFPs in gravity-matter systems are however not robust on a general basis: Generalized Proca theories provide an example of a setup in which fixed-point solutions are highly unstable and hence not robust~\cite{Heisenberg:2026rdk,Pastor-Marcos:2026nyb}.}. At the same time, preliminary evidence for a NGFP for four-Fermi theories has been recently found in~\cite{Cresswell-Hogg:2026jpe}.

The absence of a robust interacting fixed point may be surprising in light of the one-loop analysis of~\cite{Percacci:2009fh}, which found NGFPs for both the $SU(2)$ and $SU(3)$ chiral NLSMs, pointing to the possibility of their non-perturbative renormalizability. Our findings are, however, consistent with the fully non-perturbative study of the nonlinear $O(N)$ models carried out in~\cite{Flore:2012ma}. In $D=3$, their analysis of the full system finds no non-trivial fixed point; however, a NGFP exists if the four-derivative operator $(h_{ab}\partial_\mu\varphi^a\partial^\mu\varphi^b)^2$ is artificially excluded from the EFT. Since $SU(2) \simeq S^3$, this provides a useful point of contact for the $SU(2)$ model considered here, even though here the operator $h_{ab}\Box\varphi^a \Box \varphi^b$ is absent for the reasons discussed above. 

Interestingly, we observe a similar behavior: no NGFP exists in the full system in $D=4$, but when $\ell_1$, i.e., the dimensionless coupling associated with $(h_{ab}\partial_\mu\varphi^a\partial^\mu\varphi^b)^2$, is set to zero, two non-trivial fixed points emerge. This motivates us to investigate whether the disappearance of the fixed point can be understood in terms of a fixed-point collision, in which two real fixed points approach one another, eventually merging as the value of $\ell_1$ is varied. We turn to this possibility next.

For this analysis, we fix the shape function to be the exponential one: $r^{\text{Exp}}(y;1)$, having set $A$ to be $1$. We employ the same numerical strategy for finding roots as described above.

\begin{figure*}[t]
    \centering
    \includegraphics[width=0.85\columnwidth]{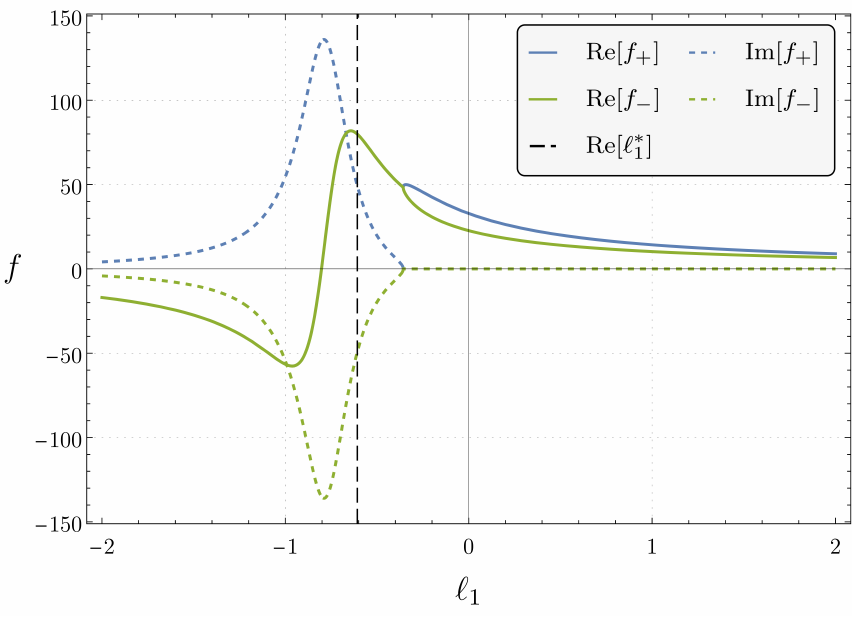}
    $\qquad$
    \includegraphics[width=0.85\columnwidth]{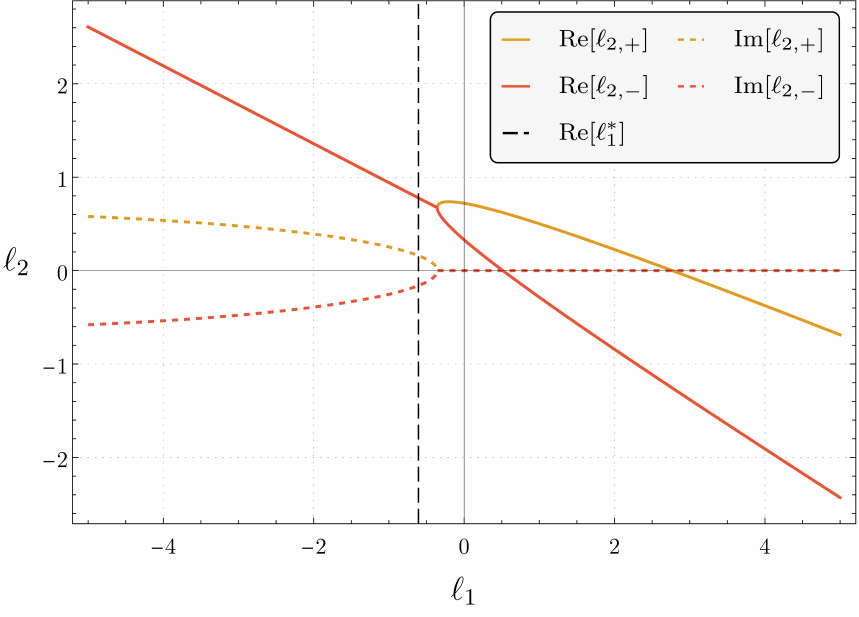}
    \caption{Partial fixed-point solutions $f(\ell_1)$ (left panel) and $\ell_2(\ell_1)$ (right panel) of $\beta_{f}$ and $\beta_{\ell_2}$. NGFPs exist when $\ell_1=0$. On the left panel, we show the two branches ${f_+}$ and ${f_-}$ (blue and green lines, respectively) for the coupling $f$. Similarly, the right panel displays the two branches $\ell_{2,+}$ and $\ell_{2,-}$ (orange and red lines, respectively) for the coupling $\ell_2$. In both panels, solid lines denote the real part of the fixed-point solutions, while a dashed colorful line depicts their imaginary parts. The black dashed vertical line represents the real part of $\ell_1^*$, where also the third beta function vanishes, $\beta_{\ell_1}=0$. This happens for ${\rm Re}[\ell_1^*] \simeq -0.607$, where both $f(\ell_1)$ and $\ell_2(\ell_1)$ acquire complex-conjugate values.}
    \label{fig:gf(g1)}
\end{figure*}

Since in $D=4$ we identified $\ell_1$ as the root cause for the absence of a NGFP, we first investigate the mechanism that makes it disappear as $\ell_1$ is slowly activated. In Fig.~\ref{fig:gf(g1)} we show the solutions of the simultaneous zeros of $\beta_{f}$ and $\beta_{\ell_2}$, with $\ell_1$ kept as free parameter. These are the curves $f(\ell_1)$ and $\ell_2(\ell_1)$, describing ``partial fixed points'', i.e., fixed points for a subset of beta functions, $\{\beta_f,\beta_{\ell_2}\}$. In solving the system, three branches of solutions exist. One is an isolated real zero, with $\ell_1=0$ and $f(0)=\ell_2(0)=0$, corresponding to the GFP. The other two branches are interacting, partial-fixed-point solutions. Fig.~\ref{fig:gf(g1)} depicts $\ell_{2 ,\pm}(\ell_1)$ (blue and green lines) and $f_{\pm}(\ell_1)$ (red and orange lines), distinguishing their real (solid lines) and imaginary (dashed lines) parts. Within this picture, the absence of a NGFP can be understood as the two real NGFPs at $\ell_1=0$ evolve as $\ell_1$ is slowly activated, and merge into a complex-conjugate pair, before the third beta function, $\beta_{\ell_1}$, can vanish. In other words, one finds no fixed point because the $f(\ell_1)$ and $\ell_2(\ell_1)$ curves are complex at the value of $\ell_1$ that also solves the last beta function $\beta_{\ell_1}[f(\ell_1^*), \ell_1^*,\ell_2(\ell_1^*)]=0$, that is $\ell_1^* \simeq -0.607 \pm 0.116 i$. 

As a complementary investigation, we study the dependence of the fixed-point structure on the spacetime dimension $D$. This is motivated by the $(2+\epsilon)$-expansion, which gives a perturbative prediction of the existence of a non-trivial fixed point for sufficiently small $\epsilon$---see e.g.~\cite{Brezin:1975sq,Friedan:1980jf,Bardeen:1976qg} for previous works in $(2+\epsilon)$-dimensions and~\cite{Higashijima:2003rp} for a study in 3 dimensions. We therefore extend our root search to generic $D$\footnote{Extending the RG analysis to generic $D$ implies that the $Q_n$ functionals also appear with negative, non-integer values of $n$; see Eq.~(5.165) of~\cite{Percacci:2017fkn} for their generalization.} and track the real solutions as the dimension is varied. In particular, we explore whether the fixed point found near two dimensions persists to larger $D$, and whether its absence in $D=4$ is associated with a collision between two real fixed points.  

\begin{figure*}[t]
    \centering
    \includegraphics[width=0.68\columnwidth]{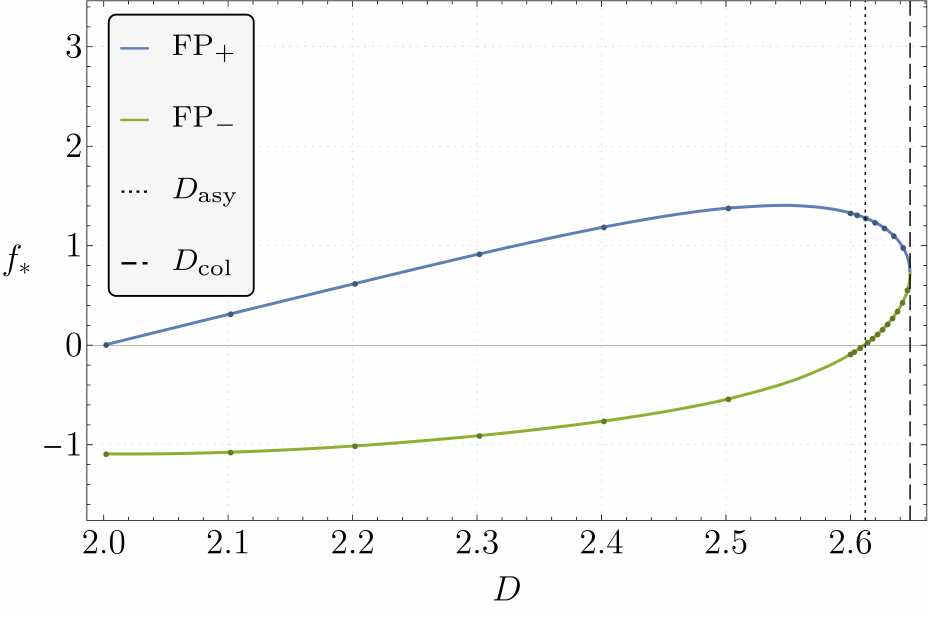}
    \hfill
    \includegraphics[width=0.68\columnwidth]{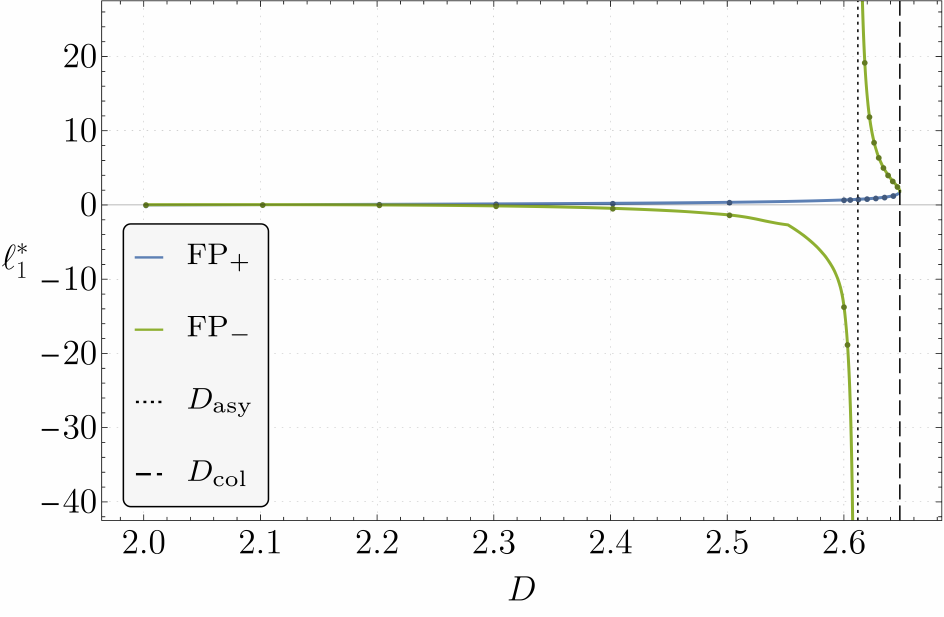}
    \hfill
    \includegraphics[width=0.68\columnwidth]{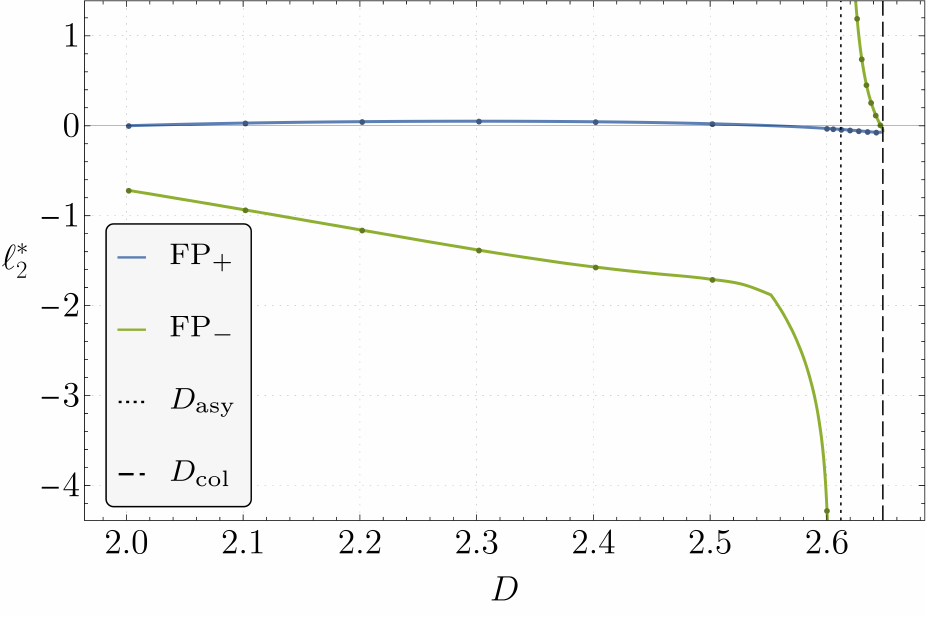}
    \caption{The two fixed-point branches (solid blue and green lines) of the $SU(2)$ beta functions as a function of $D$,
shown for $f$, $\ell_1$ and $\ell_2$. As $D$ increases, the branches approach each other
and merge. The fixed-point solutions were computed at 139 values of $D$, with a varying step-size as required by the resolution of the curves, and interpolated afterwards; for clarity, only a subset of the computed points is shown as markers. The vertical dashed lines mark $D_{\mathrm{asy}}\simeq 2.612$, where $f$ crosses zero and $\ell_1,\,\ell_2$ cease to be continuous, and $D_{\rm col}\simeq 2.648$, where the two solutions collide.}
    \label{fig:fpcoll_d}
\end{figure*}

Using the same regulator as above, and in terms of the $(f,\ell_1,\ell_2)$ couplings, we identify only two distinct real fixed-point branches, see Fig.~\ref{fig:fpcoll_d}. We label them $\rm FP_{\pm}$. As $D$ increases, the two solutions approach one another and eventually merge at a critical dimension $D_{\mathrm{col}}\simeq 2.648$, beyond which no corresponding real solutions are found. Interestingly, one of the branches extends into a region where $f<0$, displaying a suggestive similarity to the fixed-point evolution studied in~\cite{Basile:2021euh,Basile:2021krk}, in the context of stringy cosmological effective actions, where the role of $f$ is taken by the dimensionless Newton's coupling. The similarity extends to the behavior of the $\ell_1$ and $\ell_2$ coordinates: at $D_{\mathrm{asy}}\simeq 2.612$, their fixed-point branches develop an asymptotic behavior and stop being continuously connected, coinciding with a change in the character of the $f$ solution. 

To summarize, within the truncation considered, we find no robust evidence for the existence of a NGFP for the $SU(2)$ and $SU(3)$ chiral NLSM in $D=4$, thus challenging the idea that interacting fixed points should generically emerge in perturbatively non-renormalizable theories. Rather, these results raise the question of which structural features distinguish theories that admit a predictive UV completion from those that do not, and, in particular, how chiral NLSMs differ from gravity, where the existence of an interacting UV fixed point has been extensively corroborated (see~\cite{Percacci:2017fkn,Reuter:2019byg,Bonanno:2020bil,Pawlowski:2020qer,Knorr:2022dsx, Eichhorn:2022gku, Morris:2022btf, Martini:2022sll, Wetterich:2022ncl, Platania:2023srt, Saueressig:2023irs, Pawlowski:2023gym, Bonanno:2024xne,Eichhorn:2026uqj} and references therein). We turn to this in the next section.

\section{Why is asymptotic safety not generic?} \label{sec:NLSM_vs_gravity}

Beyond challenging the idea of the generic existence of a NGFP for perturbatively non-renormalizable theories, our findings motivate a closer investigation of the structural differences between chiral models and gravitational theories. In other words, \textit{if there is an analogy with gravity, where does it break down?}

Concretely, we push further the analogy proposed in~\cite{Codello:2008qq}. If the leading-order Lagrangian, Eq.~\eqref{eq:NLSM_LO}, represents the analogue of the Einstein-Hilbert term for a chiral model, then the four-derivative operators should be compared to the case of AS in quantum gravity, at quadratic order in the curvature invariants. The direct correspondence holds for $SU(2)$, which contains two four-derivative operators, leading to the identification
\begin{align}
    &R^2 \leftrightarrow (h_{ab} \partial_\mu \varphi^a \partial^\mu \varphi^b)^2 \,,\\
    &C_{\mu \nu \rho \sigma}C^{\mu \nu \rho \sigma}\leftrightarrow h_{ab} h_{cd}\partial_\mu \varphi^a \partial^\mu \varphi^c  \partial_\nu \varphi^b \partial^\nu \varphi^d \,.
\end{align}
Here, $R$ is the Ricci scalar and $C_{\mu \nu \rho \sigma}$ is the Weyl tensor. At the level of the RG flow, gravity is characterized by a NGFP~\cite{Benedetti:2009rx}. 

A comparison of the gravitational and chiral systems reveals two key differences that may account for the presence of a NGFP in the former and its absence in the latter:
\begin{itemize}
    \item The absence of a cosmological-constant term in the chiral Lagrangian. 
    \item The absence of a Faddeev-Popov contribution for non-gauge theories.
\end{itemize}
We discuss these two ingredients in detail below.

The first potential source is the cosmological constant. In gravity, this is associated with a volume term, proportional to $\sqrt{\det(g_{\mu \nu})}$, and with no derivatives. Hence, this operator contributes to the gravitational RG flow at any order in the derivative expansion. To investigate whether this may be the root cause of the absence of a NGFP for the chiral model, we studied the beta functions of~\cite{Benedetti:2009rx}, enforcing $\Lambda=0$ along the flow. Additionally, we performed a redefinition of the couplings to cast the quadratic operators into a structure similar to Eq.~\eqref{eq:chiEFT_NLO_SU2}, namely
\begin{equation}
    S_{\rm Quad}= \frac{1}{16\pi G}\int {\rm d}^4x \sqrt{g} \left( R + G_1 R^2 + G_2 C^2 \right) \,,
\end{equation}
where $C^2=C_{\mu \nu \rho \sigma}C^{\mu \nu \rho \sigma}$ and $g= \det(g_{\mu \nu})$. Interestingly, within this picture the beta functions vanish simultaneously only at the GFP, similarly to what happens for the $SU(2)$ NLSM. This suggests that the presence of a volume-term operator---specific to the geometric nature of gravity---is not just a mere quantitative difference, but that its contribution to the flow might be crucial for the existence of a NGFP. However, we stress again that setting $\Lambda=0$ \qu{by hand} is in principle inconsistent with the FRG analysis: being a zeroth-derivative operator, this term would be generated at any truncation order in the derivative expansion.

To circumvent this issue, the case of unimodular gravity seems a suitable candidate to compare to the NLSMs. In this case, the theory is partially gauge-fixed to keep $g=1$ everywhere in the spacetime manifold. Hence, $\Lambda$ does not appear in the Lagrangian as an independent operator, but rather as an integration constant in the field equations. However, to our knowledge, the full four-derivative case has not been studied yet. In~\cite{deBrito:2020xhy}, the authors studied the RG flow for the two functional dependencies $f(R)$ and $F(R_{\mu \nu}R^{\mu \nu})+R \,Z(R_{\mu \nu}R^{\mu \nu})$. Within our truncation order, these beta functions show the existence of a NGFP, in both the $f$ and $FZ$ cases. This is consistent with the existence of a NGFP for the NLSM of $SU(2)$, once one of the two next-to-leading order operators have been shut down. However, a definitive comparison would require the beta functions for a complete basis of four-derivative operators in unimodular gravity.

Independent of the specific fixed-point searches within unimodular or standard settings, the cosmological constant---being a mass term---is expected to play a role because it directly enters the inverse propagator, modifying the threshold functions with factors $(1-2\lambda)^{-n}$, which then enter the coefficients of the beta functions. Given that $\lambda<1/2$ by consistency~\cite{Percacci:2017fkn,Reuter:2019byg}, these coefficients may provide the right alternation of signs (e.g., as per Descartes' rule of signs), to guarantee the existence of at least one real root.

A second difference is that gravity is a gauge theory, so the gauge-fixing action is accompanied by a Faddeev-Popov ghost action. The ghost loop contributes to the flow equation~\eqref{eq:Wetterich_eq} with a negative sign, which enters the anomalous dimension of the Newton coupling. This helps compensate for the positive canonical term in the beta function for the dimensionless Newton's constant, which arises due to the negative mass dimension of $G$. At a qualitative level, similar negative contributions stem from the fermionic loops in the Gross-Neveu model, which may explain why a non-trivial fixed point could arise in the four-Fermi theory~\cite{Cresswell-Hogg:2026jpe}, and in gravity~\cite{Reuter:2001ag}, but not in chiral models.

\section{Conclusions} \label{sec:discussion}

In this work, we have investigated whether asymptotic safety represents a generic mechanism for the ultraviolet completion of perturbatively non-renormalizable quantum field theories, or whether the existence of interacting ultraviolet fixed points relies on specific properties of the underlying theory. To address this question, we have studied SU(2) and SU(3) chiral perturbation theories using the functional renormalization group, within a truncation of the effective action including operators up to fourth order in derivatives. These theories provide a particularly interesting testing ground, as they share important features with gravity, including nonlinear interactions and perturbative non-renormalizability, while admitting ultraviolet completions involving additional degrees of freedom.

Our analysis reveals no evidence for interacting fixed points in either the SU(2) or SU(3) models within the truncations considered. This result suggests that perturbative non-renormalizability does not, by itself, imply the existence of a non-trivial ultraviolet fixed point. While the absence of fixed points within a finite truncation cannot exclude their existence in the full theory, the consistency of our findings across the two chiral models, including within several different regularization schemes, provides a useful counterpoint to the substantial evidence supporting asymptotic safety in quantum gravity. Our work highlights that the existence of interacting fixed points akin to the Reuter fixed point is non-trivial and far from guaranteed. Complementary studies~\cite{Heisenberg:2026rdk} have provided examples that the apparent existence of an interacting fixed point does not necessarily guarantee its robustness: fixed-point values and critical exponents may exhibit a strong dependence on the choice of truncation~\cite{Heisenberg:2026rdk} and regularization scheme, in stark contrast with the stability displayed by the Reuter fixed point under similar changes~\cite{Reuter:2001ag, Lauscher:2002sq, Codello:2006in,Benedetti:2009rx,Dona:2013qba, Gies:2015tca, Gies:2016con,Biemans:2016rvp,Biemans:2017zca, Christiansen:2017cxa, Hamada:2017rvn, Knorr:2017fus, Christiansen:2017bsy, Falls:2017lst,Knorr:2018fdu,Eichhorn:2019ybe, Alkofer:2020vtb,Falls:2020qhj, Knorr:2021lll, Knorr:2021slg, Kluth:2022vnq, Pastor-Gutierrez:2022nki,Baldazzi:2023pep,Eichhorn:2025sux}. In the present work we have also investigated the disappearance of a lower-order interacting fixed point for the chiral model, showing that \emph{(i)} in $D=4$ one specific operator is the root cause for the absence of a non-trivial fixed point, meaning that its addition makes the relevant fixed point complex; \emph{(ii)} when including all operators up to four derivatives, two non-trivial fixed points exist for $D\leq 2.648$. At this critical dimension, these fixed points collide and merge into a complex-conjugate pair, which persists at higher dimensions. Interestingly, this mechanism, including the approximate value of the critical dimension, resembles what is found in certain low-energy stringy cosmological effective actions~\cite{Basile:2021euh,Basile:2021krk}.

To gain insight into the origin of this difference, we compared the structure of the chiral beta functions with those arising in gravity, at quartic order in a derivative expansion, in standard and unimodular settings. This comparison highlights two potentially relevant ingredients: the presence of a volume term in the effective action and the negative contribution stemming from the Faddeev-Popov action for gauge theories. The cosmological constant indeed acts as a mass term, which enters the propagator and the threshold functions, inducing coefficients of the form $(1-2\lambda)^{-n}$ that allow for the alternation of positive and negative signs, as $\lambda<1/2$ by consistency~\cite{Percacci:2017fkn,Reuter:2019byg}. This, in turn, may favor the development of real roots. As for the second point, Faddeev–Popov ghosts contribute with the opposite sign to bosonic fluctuations, helping to compensate the positive contributions to the beta functions associated with couplings of negative mass dimension in gravity. Such an interplay is absent in the case of chiral NLSMs. Overall, these observations suggest that the emergence of interacting fixed points may be sensitive to structural features of the theory.

Our results thus support a picture in which asymptotic safety is not an automatic consequence of perturbative non-renormalizability, but may instead depend on specific dynamical and algebraic properties of the renormalization-group flow. Establishing whether the structural differences identified here constitute general criteria for fixed-point existence will require further investigations, including extensions to higher-order truncations and comparisons with a broader class of quantum field theories, such as generic $SU(N)$ and the inclusion of the odd-parity sector through the Wess-Zumino-Witten term \cite{Wess:1971yu,Witten:1983tw}. Such studies may help clarify what makes the Reuter fixed point possible and, more generally, which properties distinguish asymptotically safe theories from those requiring alternative ultraviolet completions. 

\begin{acknowledgments}
The authors are deeply grateful to the late K. Stelle, with whom the original idea for this project was conceived. They also thank A. Eichhorn, A. Guerrieri and M. Schiffer for discussions, and G. De Brito and A. Pereira for providing the beta functions of unimodular gravity from~\cite{deBrito:2020xhy}. This research is supported by a research grant (VIL60819) from VILLUM FONDEN. The Center of Gravity is a Center of Excellence funded by the Danish National Research Foundation under grant No. 184. This project has received funding from the European Union's Horizon 2020 research and innovation programme under the Marie Sklodowska-Curie grant agreement No 101299389 and No 101131233. This work is supported by Simons Foundation International~\cite{sfi} and the Simons Foundation~\cite{sf} through Simons Foundation grant SFI-MPS-BH-00012593-11. F.F. is grateful to the Max Planck Institute for Physics in Garching, where part of the project was developed, and to Perimeter Institute for Theoretical Physics for their hospitality during the completion of this work. F.D.P. acknowledges Perimeter Institute for Theoretical Physics for their hospitality during the early stages of this work. Research at Perimeter Institute is supported in part by the Government of Canada through the Department of Innovation, Science and Economic Development Canada and by the Province of Ontario through the Ministry of Colleges and Universities. 
\end{acknowledgments}

\vfill

\bibliographystyle{apsrev4-1}
\bibliography{references.bib}

@article{Howe:1986vm,
    author = "Howe, Paul S. and Papadopoulos, G. and Stelle, K. S.",
    title = "{The Background Field Method and the Nonlinear $\sigma$ Model}",
    reportNumber = "Print-86-1354 (IAS,PRINCETON), CERN-TH-4744/87",
    doi = "10.1016/0550-3213(88)90379-3",
    journal = "Nucl. Phys. B",
    volume = "296",
    pages = "26--48",
    year = "1988"
}

@inbook{Knorr:2022dsx,
    author = "Knorr, Benjamin and Ripken, Chris and Saueressig, Frank",
    title = "{Form Factors in Asymptotically Safe Quantum Gravity}",
    eprint = "2210.16072",
    archivePrefix = "arXiv",
    primaryClass = "hep-th",
    reportNumber = "NORDITA 2022-075",
    doi = "10.1007/978-981-19-3079-9_21-1",
    bookTitle="Handbook of Quantum Gravity",
    year="2023",
    publisher="Springer Nature Singapore",
    address="Singapore",
    pages="1--49"
}

@inbook{Eichhorn:2022gku,
    author = "Eichhorn, Astrid and Schiffer, Marc",
    title = "Asymptotic Safety of Gravity with Matter",
    eprint = "2212.07456",
    archivePrefix = "arXiv",
    primaryClass = "hep-th",
    bookTitle="Handbook of Quantum Gravity",
    year="2023",
    publisher="Springer Nature Singapore",
    address="Singapore",
    pages="1--87"
}

@inbook{Morris:2022btf,
    author = "Morris, Tim R. and Stulga, Dalius",
    title = "{The Functional f(R) Approximation}",
    eprint = "2210.11356",
    archivePrefix = "arXiv",
    primaryClass = "hep-th",
    doi = "10.1007/978-981-19-3079-9_19-1",
    bookTitle="Handbook of Quantum Gravity",
    year="2023",
    publisher="Springer Nature Singapore",
    address="Singapore",
    pages="1--33"
}

@inbook{Martini:2022sll,
    author = "Martini, Riccardo and Vacca, Gian Paolo and Zanusso, Omar",
    title = "{Perturbative Approaches to Nonperturbative Quantum Gravity}",
    booktitle = "{Handbook of Quantum Gravity}",
    eprint = "2210.13910",
    archivePrefix = "arXiv",
    primaryClass = "hep-th",
    doi = "10.1007/978-981-19-3079-9_25-1",
    year="2023",
    publisher="Springer Nature Singapore",
    address="Singapore",
    pages="1--46"
}

@inbook{Wetterich:2022ncl,
    author = "Wetterich, C.",
    title = "{Quantum Gravity and Scale Symmetry in Cosmology}",
    eprint = "2211.03596",
    archivePrefix = "arXiv",
    primaryClass = "gr-qc",
    doi = "10.1007/978-981-19-3079-9_26-1",
    bookTitle="Handbook of Quantum Gravity",
    year="2023",
    publisher="Springer Nature Singapore",
    address="Singapore",
    pages="1--68"
}

@inbook{Platania:2023srt,
    author = "Platania, Alessia",
    title = "{Black Holes in Asymptotically Safe Gravity}",
    eprint = "2302.04272",
    archivePrefix = "arXiv",
    primaryClass = "gr-qc",
    reportNumber = "NORDITA 2022-085",
    doi = "10.1007/978-981-19-3079-9_24-1",
    bookTitle="Handbook of Quantum Gravity",
    year="2023",
    publisher="Springer Nature Singapore",
    address="Singapore",
    pages="1--65"
}

@article{Gies:2015tca,
    author = "Gies, Holger and Knorr, Benjamin and Lippoldt, Stefan",
    title = "{Generalized Parametrization Dependence in Quantum Gravity}",
    eprint = "1507.08859",
    archivePrefix = "arXiv",
    primaryClass = "hep-th",
    doi = "10.1103/PhysRevD.92.084020",
    journal = "Phys. Rev. D",
    volume = "92",
    number = "8",
    pages = "084020",
    year = "2015"
}

@article{Eichhorn:2025sux,
    author = "Eichhorn, Astrid and Gyftopoulos, Zois and Held, Aaron",
    title = "{Quark and lepton mixing in the asymptotically safe Standard Model}",
    eprint = "2507.18304",
    archivePrefix = "arXiv",
    primaryClass = "hep-ph",
    month = "7",
    year = "2025"
}

@article{Alkofer:2020vtb,
    author = {Alkofer, Reinhard and Eichhorn, Astrid and Held, Aaron and Nieto, Carlos M. and Percacci, Roberto and Schr{\"o}fl, Markus},
    title = "{Quark masses and mixings in minimally parameterized UV completions of the Standard Model}",
    eprint = "2003.08401",
    archivePrefix = "arXiv",
    primaryClass = "hep-ph",
    reportNumber = "Imperial/TP/2020/AH/01",
    doi = "10.1016/j.aop.2020.168282",
    journal = "Annals Phys.",
    volume = "421",
    pages = "168282",
    year = "2020"
}

@article{Eichhorn:2019ybe,
    author = "Eichhorn, Astrid and Platania, Alessia and Schiffer, Marc",
    title = "{Lorentz invariance violations in the interplay of quantum gravity with matter}",
    eprint = "1911.10066",
    archivePrefix = "arXiv",
    primaryClass = "hep-th",
    doi = "10.1103/PhysRevD.102.026007",
    journal = "Phys. Rev. D",
    volume = "102",
    number = "2",
    pages = "026007",
    year = "2020"
}

@article{Knorr:2018fdu,
    author = "Knorr, Benjamin",
    title = "{Lorentz symmetry is relevant}",
    eprint = "1810.07971",
    archivePrefix = "arXiv",
    primaryClass = "hep-th",
    doi = "10.1016/j.physletb.2019.01.070",
    journal = "Phys. Lett. B",
    volume = "792",
    pages = "142--148",
    year = "2019"
}

@article{Biemans:2017zca,
    author = "Biemans, Jorn and Platania, Alessia and Saueressig, Frank",
    title = "{Renormalization group fixed points of foliated gravity-matter systems}",
    eprint = "1702.06539",
    archivePrefix = "arXiv",
    primaryClass = "hep-th",
    doi = "10.1007/JHEP05(2017)093",
    journal = "JHEP",
    volume = "05",
    pages = "093",
    year = "2017"
}

@article{Biemans:2016rvp,
    author = "Biemans, Jorn and Platania, Alessia and Saueressig, Frank",
    title = "{Quantum gravity on foliated spacetimes: Asymptotically safe and sound}",
    eprint = "1609.04813",
    archivePrefix = "arXiv",
    primaryClass = "hep-th",
    doi = "10.1103/PhysRevD.95.086013",
    journal = "Phys. Rev. D",
    volume = "95",
    number = "8",
    pages = "086013",
    year = "2017"
}

@article{Pawlowski:2020qer,
    author = "Pawlowski, Jan M. and Reichert, Manuel",
    title = "{Quantum Gravity: A Fluctuating Point of View}",
    eprint = "2007.10353",
    archivePrefix = "arXiv",
    primaryClass = "hep-th",
    doi = "10.3389/fphy.2020.551848",
    journal = "Front. in Phys.",
    volume = "8",
    pages = "551848",
    year = "2021"
}

@article{Bonanno:2020bil,
    author = "Bonanno, Alfio and Eichhorn, Astrid and Gies, Holger and Pawlowski, Jan M. and Percacci, Roberto and Reuter, Martin and Saueressig, Frank and Vacca, Gian Paolo",
    title = "{Critical reflections on asymptotically safe gravity}",
    eprint = "2004.06810",
    archivePrefix = "arXiv",
    primaryClass = "gr-qc",
    doi = "10.3389/fphy.2020.00269",
    journal = "Front. in Phys.",
    volume = "8",
    pages = "269",
    year = "2020"
}

@article{Knorr:2021slg,
    author = "Knorr, Benjamin",
    title = "{The derivative expansion in asymptotically safe quantum gravity: general setup and quartic order}",
    eprint = "2104.11336",
    archivePrefix = "arXiv",
    primaryClass = "hep-th",
    doi = "10.21468/SciPostPhysCore.4.3.020",
    journal = "SciPost Phys. Core",
    volume = "4",
    pages = "020",
    year = "2021"
}

@article{Baldazzi:2023pep,
    author = "Baldazzi, Alessio and Falls, Kevin and Kluth, Yannick and Knorr, Benjamin",
    title = "{Robustness of the derivative expansion in asymptotic safety}",
    eprint = "2312.03831",
    archivePrefix = "arXiv",
    primaryClass = "hep-th",
    reportNumber = "NORDITA 2023-075",
    doi = "10.1103/hlrm-d4g2",
    journal = "Phys. Rev. D",
    volume = "113",
    number = "2",
    pages = "026005",
    year = "2026"
}

@article{Kluth:2022vnq,
    author = "Kluth, Yannick and Litim, Daniel F.",
    title = "{Functional renormalization for f(R{\ensuremath{\mu}}{\ensuremath{\nu}}{\ensuremath{\rho}}{\ensuremath{\sigma}}) quantum gravity}",
    eprint = "2202.10436",
    archivePrefix = "arXiv",
    primaryClass = "hep-th",
    doi = "10.1103/PhysRevD.106.106022",
    journal = "Phys. Rev. D",
    volume = "106",
    number = "10",
    pages = "106022",
    year = "2022"
}

@article{Falls:2020qhj,
    author = "Falls, Kevin and Ohta, Nobuyoshi and Percacci, Roberto",
    title = "{Towards the determination of the dimension of the critical surface in asymptotically safe gravity}",
    eprint = "2004.04126",
    archivePrefix = "arXiv",
    primaryClass = "hep-th",
    doi = "10.1016/j.physletb.2020.135773",
    journal = "Phys. Lett. B",
    volume = "810",
    pages = "135773",
    year = "2020"
}

@article{Knorr:2021lll,
    author = "Knorr, Benjamin",
    title = "{One-Loop Renormalization of Cubic Gravity in Six Dimensions}",
    eprint = "2109.09857",
    archivePrefix = "arXiv",
    primaryClass = "hep-th",
    doi = "10.1103/PhysRevLett.128.161301",
    journal = "Phys. Rev. Lett.",
    volume = "128",
    number = "16",
    pages = "161301",
    year = "2022"
}

@article{Reuter:2001ag,
    author = "Reuter, M. and Saueressig, Frank",
    title = "{Renormalization group flow of quantum gravity in the Einstein-Hilbert truncation}",
    eprint = "hep-th/0110054",
    archivePrefix = "arXiv",
    reportNumber = "MZ-TH-01-27",
    doi = "10.1103/PhysRevD.65.065016",
    journal = "Phys. Rev. D",
    volume = "65",
    pages = "065016",
    year = "2002"
}

@ARTICLE{Lauscher:2002sq,
	IDS={la02,lare},
   	author = {{Lauscher}, O. and {Reuter}, M.},
    title = "{Flow equation of quantum Einstein gravity in a higher-derivative truncation}",
  	journal = {Physical Review D},
   	eprint = {hep-th/0205062},
    year = 2002,
    month = jul,
   	volume = 66,
   	number = 2,
    eid = {025026},
    pages = {025026},
    doi = {10.1103/PhysRevD.66.025026},
   	adsurl = {http://adsabs.harvard.edu/abs/2002PhRvD..66b5026L}
}

@article{Hamada:2017rvn,
    author = "Hamada, Yuta and Yamada, Masatoshi",
    title = "{Asymptotic safety of higher derivative quantum gravity non-minimally coupled with a matter system}",
    eprint = "1703.09033",
    archivePrefix = "arXiv",
    primaryClass = "hep-th",
    doi = "10.1007/JHEP08(2017)070",
    journal = "JHEP",
    volume = "08",
    pages = "070",
    year = "2017"
}

@article{Cresswell-Hogg:2026jpe,
    author = "Cresswell-Hogg, Charlie and Litim, Daniel F.",
    title = "{Four Fermi Theory in Four Dimensions is Renormalisable}",
    eprint = "2603.19357",
    archivePrefix = "arXiv",
    primaryClass = "hep-th",
    month = "3",
    year = "2026"
}

@article{Heisenberg:2026rdk,
    author = "Heisenberg, Lavinia and Platania, Alessia and Rufrano Aliberti, Sara",
    title = "{Asymptotic safety in generalized Proca theories}",
    eprint = "2601.20944",
    archivePrefix = "arXiv",
    primaryClass = "hep-th",
    doi = "10.1016/j.physletb.2026.140546",
    journal = "Phys. Lett. B",
    volume = "878",
    pages = "140546",
    year = "2026"
}

@article{Pastor-Marcos:2026nyb,
    author = "Pastor-Marcos, Carlos and Heisenberg, Lavinia and Pastor-Guti{\'e}rrez, {\'A}lvaro and Pawlowski, Jan M. and Reichert, Manuel",
    title = "{Cosmology from asymptotically safe Proca theories}",
    eprint = "2604.01090",
    archivePrefix = "arXiv",
    primaryClass = "gr-qc",
    reportNumber = "RIKEN-iTHEMS-Report-26",
    month = "4",
    year = "2026"
}

@article{Fraaije:2022uhg,
    author = "Fraaije, Mathijs and Platania, Alessia and Saueressig, Frank",
    title = "{On the reconstruction problem in quantum gravity}",
    eprint = "2206.10626",
    archivePrefix = "arXiv",
    primaryClass = "hep-th",
    doi = "10.1016/j.physletb.2022.137399",
    journal = "Phys. Lett. B",
    volume = "834",
    pages = "137399",
    year = "2022"
}

@article{Morris:2015oca,
    author = {Morris, Tim R. and Slade, Zo{\"e} H.},
    title = "{Solutions to the reconstruction problem in asymptotic safety}",
    eprint = "1507.08657",
    archivePrefix = "arXiv",
    primaryClass = "hep-th",
    doi = "10.1007/JHEP11(2015)094",
    journal = "JHEP",
    volume = "11",
    pages = "094",
    year = "2015"
}

@article{Falls:2017lst,
    author = "Falls, Kevin and King, Callum R. and Litim, Daniel F. and Nikolakopoulos, Kostas and Rahmede, Christoph",
    title = "{Asymptotic safety of quantum gravity beyond Ricci scalars}",
    eprint = "1801.00162",
    archivePrefix = "arXiv",
    primaryClass = "hep-th",
    doi = "10.1103/PhysRevD.97.086006",
    journal = "Phys. Rev. D",
    volume = "97",
    number = "8",
    pages = "086006",
    year = "2018"
}

@article{Christiansen:2017bsy,
    author = "Christiansen, Nicolai and Falls, Kevin and Pawlowski, Jan M. and Reichert, Manuel",
    title = "{Curvature dependence of quantum gravity}",
    eprint = "1711.09259",
    archivePrefix = "arXiv",
    primaryClass = "hep-th",
    doi = "10.1103/PhysRevD.97.046007",
    journal = "Phys. Rev. D",
    volume = "97",
    number = "4",
    pages = "046007",
    year = "2018"
}

@article{Knorr:2017fus,
    author = "Knorr, Benjamin and Lippoldt, Stefan",
    title = "{Correlation functions on a curved background}",
    eprint = "1707.01397",
    archivePrefix = "arXiv",
    primaryClass = "hep-th",
    doi = "10.1103/PhysRevD.96.065020",
    journal = "Phys. Rev. D",
    volume = "96",
    number = "6",
    pages = "065020",
    year = "2017"
}

@article{Christiansen:2017cxa,
    author = "Christiansen, Nicolai and Litim, Daniel F. and Pawlowski, Jan M. and Reichert, Manuel",
    title = "{Asymptotic safety of gravity with matter}",
    eprint = "1710.04669",
    archivePrefix = "arXiv",
    primaryClass = "hep-th",
    doi = "10.1103/PhysRevD.97.106012",
    journal = "Phys. Rev. D",
    volume = "97",
    number = "10",
    pages = "106012",
    year = "2018"
}

@article{Gies:2016con,
    author = "Gies, Holger and Knorr, Benjamin and Lippoldt, Stefan and Saueressig, Frank",
    title = "{Gravitational Two-Loop Counterterm Is Asymptotically Safe}",
    eprint = "1601.01800",
    archivePrefix = "arXiv",
    primaryClass = "hep-th",
    doi = "10.1103/PhysRevLett.116.211302",
    journal = "Phys. Rev. Lett.",
    volume = "116",
    number = "21",
    pages = "211302",
    year = "2016"
}

@ARTICLE{Dona:2013qba,
	IDS={dpematter},
   	author = {{Don{\`a}}, P. and {Eichhorn}, A. and {Percacci}, R.},
    title = "{Matter matters in asymptotically safe quantum gravity}",
  	journal = {Physical Review D},
	archivePrefix = "arXiv",
   	eprint = {1311.2898},
 	primaryClass = "hep-th",
    year = 2014,
    month = apr,
   	volume = 89,
   	number = 8,
    eid = {084035},
    pages = {084035},
    doi = {10.1103/PhysRevD.89.084035},
   	adsurl = {http://adsabs.harvard.edu/abs/2014PhRvD..89h4035D}
}

@ARTICLE{Codello:2006in,
	IDS={cope,co06},
   	author = {{Codello}, A. and {Percacci}, R.},
    title = "{Fixed Points of Higher-Derivative Gravity}",
  	journal = {Physical Review Letters},
   	eprint = {hep-th/0607128},
    year = 2006,
    month = dec,
   	volume = 97,
   	number = 22,
    eid = {221301},
    pages = {221301},
    doi = {10.1103/PhysRevLett.97.221301},
   	adsurl = {http://adsabs.harvard.edu/abs/2006PhRvL..97v1301C}
}

@article{Ambjorn:2024bud,
    author = "Ambjorn, Jan and Gizbert-Studnicki, Jakub and Goerlich, Andrzej and Nemeth, Daniel",
    title = "{IR and UV Limits of CDT and Their Relations to FRG}",
    eprint = "2411.02330",
    archivePrefix = "arXiv",
    primaryClass = "hep-lat",
    doi = "10.5506/aphyspolb.55.12-a2",
    journal = "Acta Phys. Polon. B",
    volume = "55",
    pages = "12--A2",
    year = "2024"
}

@article{Ambjorn:2024qoe,
    author = "Ambj{\o}rn, Jan and Gizbert-Studnicki, Jakub and G{\H{o}}rlich, Andrzej and N{\'e}meth, D{\'a}niel",
    title = "{Is lattice quantum gravity asymptotically safe? Making contact between causal dynamical triangulations and the functional renormalization group}",
    eprint = "2408.07808",
    archivePrefix = "arXiv",
    primaryClass = "hep-lat",
    doi = "10.1103/PhysRevD.110.126006",
    journal = "Phys. Rev. D",
    volume = "110",
    number = "12",
    pages = "126006",
    year = "2024"
}

@inbook{Loll:2025eks,
    author = "Loll, R.",
    title = "{Nonperturbative quantum gravity unlocked through computation}",
    eprint = "2501.17972",
    archivePrefix = "arXiv",
    primaryClass = "hep-th",
    month = "1",
    year = "2025"
}

@article{Pastor-Gutierrez:2022nki,
    author = "Pastor-Guti{\'e}rrez, {\'A}lvaro and Pawlowski, Jan M. and Reichert, Manuel",
    title = "{The Asymptotically Safe Standard Model: From quantum gravity to dynamical chiral symmetry breaking}",
    eprint = "2207.09817",
    archivePrefix = "arXiv",
    primaryClass = "hep-th",
    doi = "10.21468/SciPostPhys.15.3.105",
    journal = "SciPost Phys.",
    volume = "15",
    number = "3",
    pages = "105",
    year = "2023"
}

@inbook{Bonanno:2024xne,
    author = "Bonanno, Alfio",
    title = "{Asymptotic Safety and Cosmology}",
    doi = "10.1007/978-981-19-3079-9_23-1",
    year = "2024"
}

@article{Eichhorn:2026uqj,
    author = "Eichhorn, Astrid",
    title = "{Asymptotically safe quantum gravity and its phenomenology -- a review}",
    eprint = "2606.21522",
    archivePrefix = "arXiv",
    primaryClass = "hep-th",
    month = "6",
    year = "2026"
}

@book{Percacci:2017fkn,
    author = "Percacci, Robert",
    title = "{An Introduction to Covariant Quantum Gravity and Asymptotic Safety}",
    doi = "10.1142/10369",
    isbn = "978-981-320-717-2, 978-981-320-719-6",
    publisher = "World Scientific",
    series = "100 Years of General Relativity",
    volume = "3",
    year = "2017"
}

@book{Reuter:2019byg,
    author = "Reuter, Martin and Saueressig, Frank",
    title = "{Quantum Gravity and the Functional Renormalization Group}: {The Road towards Asymptotic Safety}",
    isbn = "978-1-107-10732-8, 978-1-108-67074-6",
    publisher = "Cambridge University Press",
    month = "1",
    year = "2019"
}

@article{Dupuis:2020fhh,
    author = "Dupuis, N. and Canet, L. and Eichhorn, A. and Metzner, W. and Pawlowski, J.M. and Tissier, M. and Wschebor, N.",
    title = "{The nonperturbative functional renormalization group and its applications}",
    eprint = "2006.04853",
    archivePrefix = "arXiv",
    primaryClass = "cond-mat.stat-mech",
    month = "6",
    year = "2020"
}

@inbook{Saueressig:2023irs,
    author = "Saueressig, Frank",
    title = "{The Functional Renormalization Group in Quantum Gravity}",
    eprint = "2302.14152",
    archivePrefix = "arXiv",
    primaryClass = "hep-th",
    doi = "10.1007/978-981-19-3079-9_16-1",
    bookTitle="Handbook of Quantum Gravity",
    year="2023",
    publisher="Springer Nature Singapore",
    address="Singapore",
    pages="1--44"
}

@inbook{Pawlowski:2023gym,
    author = "Pawlowski, Jan M. and Reichert, Manuel",
    title = "{Quantum Gravity from Dynamical Metric Fluctuations}",
    eprint = "2309.10785",
    archivePrefix = "arXiv",
    primaryClass = "hep-th",
    bookTitle="Handbook of Quantum Gravity",
    year="2023",
    publisher="Springer Nature Singapore",
    address="Singapore",
    pages="1--70"
}

@misc{sfi,
    howpublished = "\url{ttps://www.sfi.org.bm/}",
    author = "SFI"
}

@misc{sf,
    howpublished = "\url{https://www.simonsfoundation.org/}",
    author = "SF"
}

@article{Basile:2021krk,
    author = "Basile, Ivano and Platania, Alessia",
    title = "{String tension between de Sitter vacua and curvature corrections}",
    eprint = "2103.06276",
    archivePrefix = "arXiv",
    primaryClass = "hep-th",
    doi = "10.1103/PhysRevD.104.L121901",
    journal = "Phys. Rev. D",
    volume = "104",
    number = "12",
    pages = "L121901",
    year = "2021"
}

@article{Basile:2021euh,
    author = "Basile, Ivano and Platania, Alessia",
    title = "{Cosmological {\ensuremath{\alpha}}'-corrections from the functional renormalization group}",
    eprint = "2101.02226",
    archivePrefix = "arXiv",
    primaryClass = "hep-th",
    doi = "10.1007/JHEP06(2021)045",
    journal = "JHEP",
    volume = "06",
    pages = "045",
    year = "2021"
}

@article{Codello:2008qq,
    author = "Codello, Alessandro and Percacci, Roberto",
    title = "{Fixed Points of Nonlinear Sigma Models in d{\ensuremath{>}}2}",
    eprint = "0810.0715",
    archivePrefix = "arXiv",
    primaryClass = "hep-th",
    doi = "10.1016/j.physletb.2009.01.032",
    journal = "Phys. Lett. B",
    volume = "672",
    pages = "280--283",
    year = "2009"
}

@article{Benedetti:2010nr,
    author = "Benedetti, Dario and Groh, Kai and Machado, Pedro F. and Saueressig, Frank",
    title = "{The Universal RG Machine}",
    eprint = "1012.3081",
    archivePrefix = "arXiv",
    primaryClass = "hep-th",
    reportNumber = "MZ-TH-10-42, AEI-2010-165",
    doi = "10.1007/JHEP06(2011)079",
    journal = "JHEP",
    volume = "06",
    pages = "079",
    year = "2011"
}

@article{Flore:2012ma,
    author = "Flore, Raphael and Wipf, Andreas and Zanusso, Omar",
    title = "{Functional renormalization group of the non-linear sigma model and the $O(N)$ universality class}",
    eprint = "1207.4499",
    archivePrefix = "arXiv",
    primaryClass = "hep-th",
    reportNumber = "MZ-TH-12-28",
    doi = "10.1103/PhysRevD.87.065019",
    journal = "Phys. Rev. D",
    volume = "87",
    number = "6",
    pages = "065019",
    year = "2013"
}

@article{Groh:2011dw,
    author = "Groh, Kai and Saueressig, Frank and Zanusso, Omar",
    title = "{Off-diagonal heat-kernel expansion and its application to fields with differential constraints}",
    eprint = "1112.4856",
    archivePrefix = "arXiv",
    primaryClass = "math-ph",
    reportNumber = "MZ-TH-11-43",
    month = "12",
    year = "2011"
}

@article{Percacci:2009fh,
    author = "Percacci, Roberto and Zanusso, Omar",
    title = "{One loop beta functions and fixed points in Higher Derivative Sigma Models}",
    eprint = "0910.0851",
    archivePrefix = "arXiv",
    primaryClass = "hep-th",
    reportNumber = "PI-PARTPHYS-149",
    doi = "10.1103/PhysRevD.81.065012",
    journal = "Phys. Rev. D",
    volume = "81",
    pages = "065012",
    year = "2010"
}

@article{Hasenfratz:1988rf,
    author = "Hasenfratz, P.",
    title = "{FOUR-DIMENSIONAL, ASYMPTOTICALLY FREE NONLINEAR sigma MODELS}",
    reportNumber = "BUTP-88-20-BERN",
    doi = "10.1016/0550-3213(89)90246-0",
    journal = "Nucl. Phys. B",
    volume = "321",
    pages = "139--162",
    year = "1989"
}

@article{Gasser:1983yg,
    author = "Gasser, J. and Leutwyler, H.",
    title = "{Chiral Perturbation Theory to One Loop}",
    reportNumber = "CERN-TH-3689",
    doi = "10.1016/0003-4916(84)90242-2",
    journal = "Annals Phys.",
    volume = "158",
    pages = "142",
    year = "1984"
}

@article{Weinberg:1978kz,
    author = "Weinberg, Steven",
    editor = "Deser, S.",
    title = "{Phenomenological Lagrangians}",
    reportNumber = "HUTP-78-A051A",
    doi = "10.1016/0378-4371(79)90223-1",
    journal = "Physica A",
    volume = "96",
    number = "1-2",
    pages = "327--340",
    year = "1979"
}

@article{Guerrieri:2020bto,
    author = "Guerrieri, Andrea L. and Penedones, Joao and Vieira, Pedro",
    title = "{S-matrix bootstrap for effective field theories: massless pions}",
    eprint = "2011.02802",
    archivePrefix = "arXiv",
    primaryClass = "hep-th",
    doi = "10.1007/JHEP06(2021)088",
    journal = "JHEP",
    volume = "06",
    pages = "088",
    year = "2021"
}

@article{deAzcarraga:1997ya,
    author = "de Azcarraga, J. A. and Macfarlane, A. J. and Mountain, A. J. and Perez Bueno, J. C.",
    title = "{Invariant tensors for simple groups}",
    eprint = "physics/9706006",
    archivePrefix = "arXiv",
    reportNumber = "DAMTP-97-49",
    doi = "10.1016/S0550-3213(97)00609-3",
    journal = "Nucl. Phys. B",
    volume = "510",
    pages = "657--687",
    year = "1998"
}

@article{Benedetti:2009rx,
    author = "Benedetti, Dario and Machado, Pedro F. and Saueressig, Frank",
    title = "{Asymptotic safety in higher-derivative gravity}",
    eprint = "0901.2984",
    archivePrefix = "arXiv",
    primaryClass = "hep-th",
    reportNumber = "PI-QC-115, ITP-UU-09-03, SPIN-09-03, IPHT-T09-012",
    doi = "10.1142/S0217732309031521",
    journal = "Mod. Phys. Lett. A",
    volume = "24",
    pages = "2233--2241",
    year = "2009"
}

@article{deBrito:2020xhy,
    author = "de Brito, Gustavo P. and Pereira, Antonio D. and Vieira, Arthur F.",
    title = "{Exploring new corners of asymptotically safe unimodular quantum gravity}",
    eprint = "2012.08904",
    archivePrefix = "arXiv",
    primaryClass = "hep-th",
    doi = "10.1103/PhysRevD.103.104023",
    journal = "Phys. Rev. D",
    volume = "103",
    number = "10",
    pages = "104023",
    year = "2021"
}

@article{Wetterich:1992yh,
    author = "Wetterich, Christof",
    title = "{Exact evolution equation for the effective potential}",
    eprint = "1710.05815",
    archivePrefix = "arXiv",
    primaryClass = "hep-th",
    reportNumber = "HD-THEP-92-61",
    doi = "10.1016/0370-2693(93)90726-X",
    journal = "Phys. Lett. B",
    volume = "301",
    pages = "90--94",
    year = "1993"
}

@article{Friedan:1980jf,
    author = "Friedan, D.",
    title = "{Nonlinear Models in Two Epsilon Dimensions}",
    reportNumber = "LBL-10981",
    doi = "10.1103/PhysRevLett.45.1057",
    journal = "Phys. Rev. Lett.",
    volume = "45",
    pages = "1057",
    year = "1980"
}

@article{Brezin:1975sq,
    author = "Brezin, E. and Zinn-Justin, Jean",
    title = "{Renormalization of the nonlinear sigma model in 2 + epsilon dimensions. Application to the Heisenberg ferromagnets}",
    doi = "10.1103/PhysRevLett.36.691",
    journal = "Phys. Rev. Lett.",
    volume = "36",
    pages = "691--694",
    year = "1976"
}

@article{Bardeen:1976qg,
    author = "Bardeen, William A. and Bander, Myron",
    title = "{Phase Transition in an O(N) Gauge Model in Two-Dimensions}",
    reportNumber = "FERMILAB-PUB-76-038-THY, FERMILAB-PUB-76-038-T",
    doi = "10.1103/PhysRevD.14.2117",
    journal = "Phys. Rev. D",
    volume = "14",
    pages = "2117",
    year = "1976"
}

@article{Higashijima:2003rp,
    author = "Higashijima, Kiyoshi and Itou, Etsuko",
    title = "{Three-dimensional nonlinear sigma models in the Wilsonian renormalization method}",
    eprint = "hep-th/0304194",
    archivePrefix = "arXiv",
    reportNumber = "OU-HET-439",
    doi = "10.1143/PTP.110.563",
    journal = "Prog. Theor. Phys.",
    volume = "110",
    pages = "563--578",
    year = "2003"
}

@article{Wess:1971yu,
    author = "Wess, J. and Zumino, B.",
    title = "{Consequences of anomalous Ward identities}",
    doi = "10.1016/0370-2693(71)90582-X",
    journal = "Phys. Lett. B",
    volume = "37",
    pages = "95--97",
    year = "1971"
}

@article{Witten:1983tw,
    author = "Witten, Edward",
    title = "{Global Aspects of Current Algebra}",
    reportNumber = "PRINT-83-0262 (PRINCETON)",
    doi = "10.1016/0550-3213(83)90063-9",
    journal = "Nucl. Phys. B",
    volume = "223",
    pages = "422--432",
    year = "1983"
}

\let\clearpage\relax 
\onecolumngrid

\appendix

\section{Conventions and useful identities}\label{Appendix:0}
In this appendix we report the convention used along the computations together with some useful identities in the case $G=SU(3)$. Let us start with the Riemann tensor. Given a vector $V^a$ in the tangent space we use the convention
\begin{equation}
     [\nabla_a, \nabla_b]V^c\equiv R^c \,_{d a b}V^d \,.
\end{equation}

For a NLSM, two covariant derivatives exist: $\nabla_a$, acting on the target manifold $\mathcal N$ and compatible with $h_{ab}$, and $\nabla_\mu$ acting on the spacetime manifold $\mathcal M$. The fields $\varphi^a$ are scalar functions on $\mathcal M$, so that $\nabla_\mu \varphi^a=\partial_\mu \varphi^a$. However, vectors on the tangent space of $\mathcal N$ have non-vanishing covariant derivative $\nabla_\mu$,
\begin{equation}
    \nabla_\mu V^a(\varphi)\equiv\partial_\mu \varphi^b \nabla_b V^a(\varphi) \,,
\end{equation}
and similarly for the other tensors on $\mathcal M$. In particular $\nabla_\mu h_{ab}=0$ by definition. Note that, even if the spacetime curvature is flat, one can have non-vanishing curvature in the field space, given by~\cite{Percacci:2009fh}
\begin{align} \label{eq:field_strength}
 F_{\mu \nu} \,^a\,_b V^b:= [\nabla_\mu , \nabla_\nu] V^a = \partial_\mu  \varphi^c \partial_\nu  \varphi^d R_{cd}\,^a \,_b V^b \implies  F_{\mu \nu} \,^a\,_b= \partial_\mu \varphi^c \partial_\nu  \varphi^d R_{cd}\,^a \,_b\,.
\end{align}

Another identity comes from considering a geodesic $u^a$, and computing the commutator with $\nabla_\mu$
\begin{align} \label{eq:geodesic_comm}
   [u^a \nabla_a,  \nabla_\mu]V^c=u^a \partial_\mu \Phi^b [\nabla_a, \nabla_b]V^c+ u^a \nabla_a \partial_\mu \Phi^c- \partial_\mu \Phi^a \nabla_a u^c= u^a \partial_\mu \Phi^b R^c \,_{dab}V^d\,.
\end{align}
This is particularly useful when expanding the NLSM around a background $\bar \varphi^a$ along a geodesic.

\subsection{Riemann and Ricci tensors for $SU(2)$ and $SU(3)$}
As explained in~\cite{Codello:2008qq}, $SU(2) \simeq S^3$ is a maximally symmetric manifold. In this case, both the Riemann and Ricci tensor can be expressed through the metric $h_{ab}$ and $R$ is constant. Within the conventions used in this work, we have
\begin{equation} \label{eq:max_symm}
    R_{abcd}=  h_{ac} h_{bd}- h_{ad}h_{bc}\,, \qquad R_{ab}= 2  h_{ab}  \,, \qquad G=SU(2) \,.
\end{equation}

The case of $SU(3)$ is less symmetric. It is not anymore a maximally symmetric space, but it remains an Einstein manifold. In other words the Ricci tensor is still proportional to the metric, that is
\begin{equation} \label{eq:Einstein}
    R_{ab}=  \frac{3}{4}h_{ab}  \,, \qquad G=SU(3) \,.
\end{equation}
The Riemann tensor is a function of the structure constants~\cite{Percacci:2009fh}. Let $T_a$ be the generators of the Lie algebra of $SU(3)$, we have
\begin{equation}
    [T_a,T_b]=:if_{abc} T_c, \qquad  R_{abcd}= \frac{1}{4}f_{abe}f^e \,_{cd}  \,, \qquad G=SU(3) \,. \label{eq:generators_and_Riem}
\end{equation} 
These relations are useful to expand and simplify the modified inverse propagator on the right-hand side of the Wetterich equation, Eq.~\eqref{eq:Wetterich_eq}.

\subsection{Useful identities for SU(3)}
Finally, we collect some useful identities for the Riemann tensor of $SU(3)$, which enter the computation of the RG flow. In particular, we focus on contractions of pairs of Riemann tensors, which arise on the right-hand side of the Wetterich equation. We can define, as in~\cite{deAzcarraga:1997ya}, the matrix
\begin{equation}
   (\mathcal F_a)_{bc}:=f_{bac}\,,
\end{equation}
and the tensor
\begin{equation}
   \mathcal T_{abcd}^{(3)}:=\frac{1}{2}(f_{ab}\,^e f_{cde}+f_{ad}\,^e f_{cbe})\,.
\end{equation}
Using the symmetries of the Riemann tensor, we can thus express $\mathcal T_{abcd}^{(3)}$ entirely in terms of the curvature tensor
\begin{equation}
   \mathcal T_{abcd}^{(3)}-\mathcal T_{acbd}^{(3)}=6 R_{adbc} \iff \mathcal T_{abcd}^{(3)}=2(R_{abdc}+R_{adbc})\,.
\end{equation}
Taking the Bianchi identities into account, one can construct four independent quadratic contractions of the Riemann tensor: $R_{ab}\,^{cd}R_{lmcd}$, $R_{a}\,^{c}\,_b \,^dR_{lcmd}$, $R_{a}\,^{c}\,_b \,^dR_{ldmc}$, and $R_{abcd}R_{l}\,^c\,_m\,^d$. The first identity is then given by summing over the last two indices of two Riemann tensors,
\begin{equation} \label{eq:Id1}
R_{ab}\,^{cd}R_{lmcd}= \frac{1}{16}f_{abe}f^{ecd}f_{lmk}f^k \,_{cd}= \frac{1}{16}f_{abe}f_{lmk}{\rm Tr}(\mathcal F^k F^e)= \frac{3}{4}R_{ablm}\,,
\end{equation}
where in the last step we have used the fact that ${\rm Tr}(\mathcal F^k F^e)=3 \delta^{ek}$ as proven in~\cite{deAzcarraga:1997ya}. Similarly, one can compute, for the contraction of the second and fourth indices
\begin{align} \label{eq:Id2}
   R_{a}\,^{c}\,_b \,^dR_{lcmd}&= \frac{1}{16} {\rm Tr}( \mathcal F_a  \mathcal F_l  \mathcal F_m  \mathcal F_b) \nonumber\\
   &=\frac{1}{16} \left(- \mathcal T_{amlb}^{(3)} + \frac{3}{4}(\delta_{am}\delta_{lb}+\delta_{al}\delta_{mb}+\delta_{ab}\delta_{lm}) \right) \\
   &=\frac{1}{16} \left(-2(R_{almb}+R_{abml})  + \frac{3}{4}(\delta_{am}\delta_{lb}+\delta_{al}\delta_{mb}+\delta_{ab}\delta_{lm}) \right) \nonumber\,,
\end{align}
and
\begin{align} \label{eq:Id3}
   R_{a}\,^{c}\,_b \,^dR_{ldmc}&= \frac{1}{16} {\rm Tr}( \mathcal F_a  \mathcal F_m  \mathcal F_l  \mathcal F_b) \nonumber\\
   &=\frac{1}{16} \left(- \mathcal T_{almb}^{(3)} + \frac{3}{4}(\delta_{am}\delta_{lb}+\delta_{al}\delta_{mb}+\delta_{ab}\delta_{lm}) \right)\\
   &=\frac{1}{16} \left(-2(R_{amlb}+R_{ablm})  + \frac{3}{4}(\delta_{am}\delta_{lb}+\delta_{al}\delta_{mb}+\delta_{ab}\delta_{lm}) \right) \nonumber\,.
\end{align}
From Eq.s~\eqref{eq:Id2} and~\eqref{eq:Id3}, one can compute one last identity. Indeed, using the Bianchi identity, we have
\begin{align} \label{eq:Id4}
   R_{abcd}R_{l}\,^c\,_m\,^d&=-R_{adbe}R_{l}\,^c\,_m\,^d+ R_{acbd}R_{l}\,^c\,_m\,^d= \nonumber\\
   &=\frac{1}{16} \left({\rm Tr}( \mathcal F_a  \mathcal F_l  \mathcal F_m  \mathcal F_b)-{\rm Tr}( \mathcal F_a  \mathcal F_m  \mathcal F_l  \mathcal F_b) \right)= \frac{3}{8}R_{ablm}\,.
\end{align}
As a consistency check, Eq.~\eqref{eq:Id4} can also be computed by applying the Bianchi identity to the other Riemann tensor
\begin{align}
   R_{ab}\,^{cd}R_{lcmd}&=-R_{ab}\,^{dc}R_{ldmc}+ R_{ab}\,^{cd}R_{lmcd}\implies 2R_{ab}\,^{cd}R_{lcmd}=\frac{3}{4}R_{ablm}\,.
\end{align}
This concludes the set of identities and relations that we used to expand and simplify the right-hand side of the RG flow equation, which we partially develop in the next appendix.

\section{Regularized inverse propagator for NLSM}\label{Appendix:1}
In this appendix we report the computation for the regularized propagator of the NLSM. The Lagrangian we consider can be ordered in powers of derivatives,
\begin{equation}
    \mathcal L= \mathcal L_{\rm LO}+ \mathcal L_1 + \mathcal L_2 + \mathcal L_3\,,
\end{equation}
with
\begin{align*}
    &\mathcal L_{\rm LO}=\frac{1}{2 F}  h_{ab}(\varphi) \partial_{\mu}\varphi^a\partial^{\mu}\varphi^b \,, \qquad\mathcal L_1=-\frac{L_1}{4F}   (h_{ab} \partial_\mu \varphi^a \partial^\mu \varphi^b)^2 \,,\\
   & \mathcal L_2= -\frac{L_2}{4 F}h_{ab} h_{cd}\partial_\mu \varphi^a \partial^\mu \varphi^c  \partial_\nu \varphi^b \partial^\nu \varphi^d \,, \qquad\mathcal L_3= -\frac{L_3}{4F}   R_{a b c d} \partial_{\mu} \varphi^{a}\partial_{\nu} \varphi^{b}\partial^{\mu} \varphi^{c}\partial^{\nu} \varphi^{d}  \,,
\end{align*}
and $\ell_3=0$ for $G=SU(2)$. To expand in perturbations $\xi^a$ with the exponential map~\eqref{eq:exp_map}, we follow~\cite{Howe:1986vm}. We consider a geodesic $\Phi^a(s;x)$ on $\mathcal{N}$, parametrized by $s\in[0,1]$, such that
\begin{align}
    \Phi^a(0;x)= \bar \varphi^a(x) \,, \qquad \Phi^a(1;x)=  \varphi^a(x) \,, \qquad \frac{{\rm d}\Phi^a(s;x)}{{\rm d}s} \biggr|_{s=0}=\xi^a(x) \,.
\end{align}
The fluctuation field $\xi^a$ parametrizes deviations from $\bar \varphi$ and thus plays the role of the quantum fluctuations. Having promoted the Lagrangian to $\mathcal L(s):=\mathcal L[\Phi(s)]$, it can be expanded by
\begin{align}
   \mathcal L[\varphi]=\mathcal L(1) = \sum_{n=0}^\infty \frac{1}{n!}\frac{{{\rm{d}}^n}}{{\rm d}s^n} \mathcal L[\Phi(s)] \biggr|_{s=0}=\mathcal L(0)+\frac{{\rm d}\mathcal L}{{\rm d}s}(0)+\frac{1}{2}\frac{{\rm d}^2\mathcal L}{{\rm d}s^2}(0) + \cdots\,.
\end{align}
The expansion in $s$ around $s=0$ corresponds to an expansion in $\xi^a$. Therefore, the quadratic contribution is determined by the third term in the sum above. At the technical level, we consider the tangent vector $u^a \nabla_a:= {\rm d}/{\rm d}s$ and we make repeated use of identities like those in Eq.~\ref{eq:geodesic_comm}.

For $G=SU(2)$, the covariant expansion gives the quadratic part of the Lagrangian
\begin{align} \label{eq:quad_L}
    \mathcal{L}^{(2)}= \frac{h_{ab} \nabla_\mu \xi^a \nabla^\mu \xi ^b}{2F}   +\frac{ \xi^a \xi^b R_{cbad}}{2F } \mathcal U^{cd}(\bar \varphi)+\frac{ \nabla_\mu \xi^a \nabla_\nu \xi^b \partial_\rho \bar \varphi^c \partial_\sigma \bar \varphi^d}{2F } \mathcal T^{\mu \nu \rho \sigma}_{abcd}(\bar \varphi)
\end{align}
where $h_{ab}=h_{ab}(\bar \varphi)$, $R_{abcd}=R_{abcd}(\bar \varphi)$ are understood and
\begin{align}
     \mathcal U^{cd}(\bar \varphi)=  \partial_\mu \bar \varphi^c \partial^\mu \bar \varphi^d \left( 1 - L_1 \partial_\nu \bar \varphi^l \partial^\nu \bar \varphi^m h_{lm} \right)- L_2 \partial_\mu\varphi^c \partial_\nu \bar \varphi^d  \partial^\mu \bar \varphi^m \partial^\nu \bar \varphi^l h_{lm} \,, \quad (G=SU(2)) \,, \label{eq:U_tensor}
\end{align}
together with 
\begin{align}
     \mathcal T^{\mu \nu \rho \sigma}_{abcd}(\bar \varphi)= - h_{ab}h_{cd}(L_1 \eta^{\rho \sigma} \eta^{\mu \nu}+ L_2 \eta^{\mu \rho } \eta^{\sigma \nu})- h_{ac} h_{bd}(L_2 \eta^{\rho \sigma} \eta^{\mu \nu}+2 L_1 \eta^{\mu \rho } \eta^{\sigma \nu}+L_2 \eta^{\nu \rho } \eta^{\mu \sigma}) , \quad (G=SU(2)) \,. \label{eq:T_tensor}
\end{align}
For $G=SU(3)$, the quadratic Lagrangian contains also the term proportional to $L_3$,
\begin{align}
    \mathcal L^{(2)}_3=& - \frac{L_3}{F} \biggl(\xi^a \xi^b R_{cbad} R_{nm}\,^d\,_l \partial_\mu \varphi^n \partial^\mu \varphi^c \partial_\nu \varphi^m \partial^\nu \varphi^l + R_{acbd} \nabla_\mu \xi^a \nabla^\mu \xi^b \partial_\nu \varphi^c \partial^\nu \varphi^d \nonumber \\
        & + R_{acbd} \nabla_\mu \xi^a \partial^\mu \varphi^b \nabla_\nu \xi^c \partial^\nu \varphi^d +R_{acbd} \nabla_\mu \xi^a \partial^\mu \varphi^b \partial^\nu \varphi^c \nabla_\nu \xi^d \biggr) \,,
\end{align}
from which one can directly read the modification to the two tensors $\mathcal T$ and $\mathcal U$ defined as in Eq.~\eqref{eq:quad_L}. In both cases, after integration by parts, one gets
\begin{align}
    \mathcal{L}^{(2)} &= \frac{1}{2 F} \xi^a \left( - h_{ab} \square+ R_{cbad}  \mathcal U^{cd} - \nabla_\mu \mathcal T^{\mu \nu \rho \sigma}_{abcd}\partial_\rho \bar \varphi^c \partial_\sigma \bar \varphi^d \nabla_\nu \right) \xi^b \\ &= \frac{1}{2 F} \xi^a \left( - h_{ab} \square+ \Sigma_{ab} \right) \xi^b \nonumber \,,
\end{align}
where $\square=\nabla_\mu\nabla^\mu$ and the leftmost derivative in the first line acts on all quantities to its right. Following~\cite{Flore:2012ma}, we split the Hessian as
\begin{align} \label{eq:Sigma_appendix}
    \Sigma_{ab}(\bar \varphi)  =   \mathcal B^{\mu \nu}_{ab} \nabla_\mu \nabla_\nu + \mathcal C^\nu_{ab} \nabla_\nu + \mathcal D_{ab}  \,,
\end{align}
with
\begin{align}
    \mathcal  B^{\mu \nu}_{ab}= -  \mathcal T^{\mu \nu \rho \sigma}_{abcd}\partial_\rho \bar \varphi^c \partial_\sigma \bar \varphi^d \,,  \quad \mathcal C^\nu_{ab} = \nabla_\mu \mathcal  B^{\mu \nu}_{ab} \,, \quad \mathcal D_{ab}=R_{cbad}  \mathcal U^{cd} \,.
\end{align}
We note that $\nabla_\mu\mathcal T^{\mu \nu \rho \sigma}_{abcd}=0$, therefore the matrix $C^\mu_{ab}$ contains at least two fields $\bar \varphi$ and three derivatives of the form $\mathcal C \sim \partial^2 \bar \varphi \partial \bar \varphi$. To regularize the propagator, we employ the regulator~\eqref{eq:regulator}, which amounts to the replacement
\begin{equation} \label{eq:regularization}
   -h_{ab}\square \to \mathcal P_{k,ab}(-\square)= h_{ab}P_k(-\square) \,, \qquad P_k(-\square)=-\square+R_k(-\square)  \,.
\end{equation}
This concludes the computation of the regularized inverse propagator, $(\Gamma_k^{(2)}+\mathcal{R}_k)$. 

\section{Heat-kernel for non-minimal differential operator}\label{Appendix:2}
The right-hand side of Eq.~\eqref{eq:Wetterich_eq} can be computed through the standard Parker-Fulling expansion. Starting from Eq.~\eqref{eq:Hessian} regularized as~\eqref{eq:regularization} we have~\cite{Benedetti:2010nr}
\begin{align}
{\rm Tr} \left( (\Gamma_k^{(2)}+ \mathcal R_k)^{-1}  k \partial_k \mathcal R_k \right)=&{\rm Tr} \left( \mathcal P_k^{-1}   k \partial_k \mathcal R_k \right)-{\rm Tr} \left( \Sigma \mathcal P_k^{-1}   (k \partial_k \mathcal R_k) \mathcal P_k^{-1} \right)+{\rm Tr} \left( \Sigma^2 (\mathcal P_k^{-1} )^3  (k \partial_k \mathcal R_k)  \right) + \mathcal{O}(\Sigma^3) \,, \label{eq:rhs_Wetterich}
\end{align}
where matrix multiplication is understood. In the term linear in $\Sigma$, we have used the cyclicity of the trace. For the term quadratic in $\Sigma$, the operators can be freely commuted, since any non-vanishing commutator would contribute only beyond the order retained in our truncation~\cite{Flore:2012ma,Knorr:2021slg}. 

The evaluation of the supertrace requires heat-kernel techniques for non-minimal differential operators
\begin{align} \label{eq:f_n}
{\rm Tr} \left( \nabla_{\mu_1} \dots \nabla_{\mu_n}f_n(\square) \right) \quad \mbox{where} \quad f_n(\square)= ( P_k^{-1}(\square) )^n  (k \partial_k  R_k(\square)) \,.
\end{align}
We evaluate these traces using the off-diagonal heat-kernel expansion developed in~\cite{Groh:2011dw,Flore:2012ma}. In $D$ spacetime dimensions, introducing the inverse Laplace transform of $f_n$,
\begin{align}
f_n(\square)=\int_0^\infty {\rm d}s \,L^{-1}[f_n](s)\,e^{-s\square}\,,
\end{align}
the traces can be expanded for small $s$ as
\begin{align}
 {\rm Tr}(\nabla_{\mu_1} \dots \nabla_{\mu_n} e^{-s \square}) = \sum_{j=0}^\infty \frac{B_{\mu_1 \dots \mu_n,j}}{(4 \pi s)^{D/2}} s^j   \,,
\end{align}
where the heat-kernel coefficients $B_{\mu_1\cdots\mu_n,j}$ were computed in~\cite{Groh:2011dw}. We are interested in computing a few of these coefficients. The one with a single derivative is
\begin{align} \label{eq:tr_1der}
{\rm Tr} \left( \nabla_{\mu} f_n(\square) \right)= \int_0^\infty {\rm d}s \, L^{-1}[f_n](s) {\rm Tr}(\nabla_{\mu} e^{-s \square}) = \sum_{j=0}^\infty \int_0^\infty {\rm d}s \, L^{-1}[f_n](s)\frac{s^j}{(4 \pi s)^{D/2}} B_{\mu,j}  \,,
\end{align}
where $B_\mu$, given in~\cite{Groh:2011dw}, is
\begin{align}
 B_{\mu,0}= 0 \,, \qquad B_{\mu,1} =- \frac{1}{6} {\rm tr} \int {\rm d}^Dx \nabla^\nu F_{\nu \mu}\,^a \,_b  \,,
\end{align}
and $F_{\mu \nu ab}$ is defined in~\eqref{eq:field_strength}. Therefore, at the lowest order, ${\rm Tr} \left( \nabla_{\mu} f_n(\square) \right)$ contributes with at least two background fields and two derivatives. In particular, $B_{\mu,0}= 0$ and, as a consequence, the operator $\mathcal C^\mu_{ab}$ in Eq.~\eqref{eq:Sigma} does not contribute to the flow. Indeed,  ${\rm Tr} \left( \nabla_{\mu} f_n(\square) \right)$ contributes with at least two background fields and two background derivatives, whereas $\mathcal C^\mu\sim \partial^2\bar\varphi\,\partial\bar\varphi$. Their combination hence exceeds the chosen truncation. This observation agrees with the results of~\cite{Percacci:2009fh,Flore:2012ma}. In a similar manner, one has
\begin{align} \label{eq:tr_2der}
{\rm Tr} \left( \nabla_{(\mu} \nabla_{\nu)} f_n(\square) \right) = \sum_{j=-1}^\infty \int_0^\infty {\rm d}s \, L^{-1}[f_n](s)\frac{s^{j}}{(4 \pi s)^{D/2}} B_{\mu \nu,j}  \,,
\end{align}
with
\begin{align} 
B_{\mu \nu,-1}=- \frac{1}{2} \int {\rm d}^4 x \, \eta_{\mu \nu} \,, \quad B_{\mu \nu,0}=0 \,, 
\end{align}
while $B_{\mu \nu,1} \sim (F_{\mu \nu})^2 \sim (\partial \bar \varphi)^4$ and does not contribute to our truncation. To get the full contribution of the term with two derivatives, we note that~\cite{Groh:2011dw}
\begin{align} 
{\rm Tr} \left( \nabla_{\mu} \nabla_{\nu} f_n(\square) \right) = {\rm Tr} \left( \nabla_{(\mu} \nabla_{\nu)} f_n(\square) \right) + {\rm Tr} \left( \nabla_{[\mu} \nabla_{\nu]} f_n(\square) \right)  \,,
\end{align}
and that (we recall here that $2\nabla_{[\mu} \nabla_{\nu]}:= [\nabla_\mu, \nabla_\nu]$)
\begin{align} 
 {\rm Tr} \left( \nabla_{[\mu} \nabla_{\nu]} f_n(\square) \right)=\frac 12 \int_0^\infty {\rm d}s \, L^{-1}[f_n](s)\frac{1}{(4 \pi s)^{D/2}}  {\rm tr} \int {\rm d}^Dx F_{\mu \nu}\,^a \,_b + \mathcal{O}(F)  \,.
\end{align}
Along the same lines, one can show that ${\rm Tr} \left( \nabla_{\mu} \nabla_{\nu} \nabla_{\rho} e^{-s \square}\right)$ does not contribute within our truncation. The last term that can be generated by~\eqref{eq:tr_2der} is
\begin{equation}
  {\rm Tr} \left( \nabla_{\mu} \nabla_{\nu} \nabla_{\rho} \nabla_{\sigma}  e^{-s \square}\right)  \,. 
\end{equation}
Expanding again for small $s$
\begin{equation}
  {\rm Tr} \left( \nabla_{(\mu} \nabla_{\nu} \nabla_{\rho} \nabla_{\sigma)}  e^{-s \square}\right)=\sum_{j=-2}^\infty \int_0^\infty {\rm d}s \, L^{-1}[f_n](s)\frac{s^{j}}{(4 \pi s)^{D/2}} B_{\mu \nu\rho \sigma,j}  \,, 
\end{equation}
one sees that the only Heat Kernel coefficient contributing to this trace term is
\begin{equation}
  B_{\mu \nu\rho \sigma,-2} =\frac{3}{4} {\rm tr}\int {\rm d}^D x \, \eta_{(\mu \nu} \eta_{\rho \sigma)}\,, 
\end{equation}
while the other ones can be ignored. 

\section{Full chiral beta functions}\label{Appendix:4}
In this appendix, we report the beta functions of the dimensionful chiral couplings in $D=4$, written in terms of the $Q_{l,n}$ functionals, which also contain the anomalous dimension induced by the $1/F$ prefactor of the regulator. The $Q_{l,n}$ functionals are defined as~\cite{Flore:2012ma}
\begin{equation}
    Q_{l,n}= \frac{1}{(4 \pi)^2} \int_0^\infty s^{-l}L^{-1}[f_n](s)  \,,
\end{equation}
where the operator $L^{-1}[\cdot]$ is the inverse Laplace transform and the functions $f_n$ are defined in Eq.~\eqref{eq:f_n}.

For the $SU(2)$ case, the beta functions are
\begin{align}
\beta_{F} &= -F^{2}\left(2 Q_{2,2} + 7 L_{1} Q_{3,2} + 4 L_{2} Q_{3,2}\right)\,, \\[6pt]
\beta_{L_1} &= \frac{1}{3} F \Big( Q_{0,1} - 3\Big[ 2 Q_{2,3} + 9 L_{2}^{2} Q_{4,3}
  + L_{1}\left(7 L_{1} Q_{3,2} + 20 Q_{3,3} + 50 L_{1} Q_{4,3}\right) \nonumber \\
&\qquad\qquad + L_{2}\left(-Q_{2,2} + 4 L_{1} Q_{3,2} + 14 Q_{3,3} + 52 L_{1} Q_{4,3}\right) \Big] \Big)\,, \\[6pt]
\beta_{L_2} &= -\frac{1}{3} F \Big( Q_{0,1} + 3\Big[ 2 Q_{2,3} - L_{2}\left(Q_{2,2} + 10 Q_{3,3}\right)
  + 4 L_{1}^{2} Q_{4,3} + 4 L_{2}^{2}\left(Q_{3,2} + 6 Q_{4,3}\right) \nonumber \\
&\qquad\qquad + L_{1}\left(-2 Q_{2,2} + 7 L_{2} Q_{3,2} - 4 Q_{3,3} + 20 L_{2} Q_{4,3}\right) \Big] \Big)\,.
\end{align}

For $SU(3)$, they read 
\begin{align}
\beta_{F} &= -\frac{1}{8} F^{2}\Big(6 Q_{2,2} + \left(136 L_{1} + 52 L_{2} + 9 L_{3}\right) Q_{3,2}\Big)\,, \\[6pt]
\beta_{L_1} &= -\frac{1}{128} F \Big( 12 Q_{2,3} + 192 L_{2} Q_{3,3} + 36 L_{3} Q_{3,3} + 1472 L_{2}^{2} Q_{4,3}
  + 480 L_{2} L_{3} Q_{4,3} + 45 L_{3}^{2} Q_{4,3} + 128 L_{1}^{2}\left(17 Q_{3,2} + 110 Q_{4,3}\right) \nonumber \\
&\qquad - 16 L_{1}\left(6 Q_{2,2} - 52 L_{2} Q_{3,2} - 9 L_{3} Q_{3,2} - 48 Q_{3,3} - 656 L_{2} Q_{4,3} - 108 L_{3} Q_{4,3}\right) \Big)\,, \\[6pt]
\beta_{L_2} &= -\frac{1}{64} F \Big( 12 Q_{2,3} + 36 L_{3} Q_{3,3} + 256 L_{1}^{2} Q_{4,3} + 45 L_{3}^{2} Q_{4,3}
  + 32 L_{2}^{2}\left(13 Q_{3,2} + 58 Q_{4,3}\right) \nonumber \\
&\qquad - 8 L_{2}\Big[6 Q_{2,2} - 9 L_{3} Q_{3,2} + 12 L_{3} Q_{4,3} - 8 L_{1}\left(17 Q_{3,2} + 20 Q_{4,3}\right)\Big] \Big)\,, \\[6pt]
\beta_{L_3} &= \frac{1}{8} F \Big( Q_{0,1} + 8 L_{2} Q_{2,2} - 3 L_{3} Q_{2,2} - 2 Q_{2,3} - 52 L_{2} L_{3} Q_{3,2}
  - 9 L_{3}^{2} Q_{3,2} - 80 L_{2} Q_{3,3} \nonumber \\
&\qquad - 6 L_{3} Q_{3,3} - 192 L_{2} L_{3} Q_{4,3} - 6 L_{3}^{2} Q_{4,3}
  - 8 L_{1}\left(2 Q_{2,2} + 17 L_{3} Q_{3,2} + 4 Q_{3,3} + 12 L_{3} Q_{4,3}\right) \Big)\,.
\end{align}
We use these beta functions in Sect.~\ref{sec:NO_FP} to investigate the existence of non-trivial fixed points in chiral NLSM.

\vfill

\end{document}